\documentclass[pra,preprint]{revtex4}
\usepackage{graphics}
\usepackage{amsfonts}
\usepackage{bm}

\begin{document}
\title{
SELF-CONSISTENT SEPARABLE RPA APPROACH FOR SKYRME FORCES: AXIAL NUCLEI}
\author{V.O. Nesterenko$^{1}$, J. Kvasil $^{2}$, W. Kleinig $^{1,3}$, P.-G. Reinhard $^{4}$,
and D.S. Dolci $^{1}$}
\address{$^{1}$ Bogoliubov Laboratory of Theoretical Physics,
Joint Institute for Nuclear Research, Dubna, Moscow region, 141980, Russia}
\address{$^{2}$ Institute of Particle and Nuclear Physics, Charles University, CZ-18000 Praha 8,
Czech Republic}
\address{$^{4}$  Technische Univirsitat Dresden, Inst. f\"ur Analysis,
   D-01062, Dresden, Germany}
\address{$^{3}$Institut f\"ur Theoretische Physik II,
Universit\"at Erlangen, D-91058, Erlangen, Germany}
\begin{abstract}
{\sl\small The self-consistent separable RPA (random phase
approximation) method is formulated for Skyrme forces with pairing.
The method is based on a general self-consistent procedure for factorization of
the two-body interaction. It is relevant for various density- and
current-dependent functionals. The contributions of the time-even and
time-odd Skyrme terms as well as of the Coulomb and pairing terms to the
residual interaction are  taken self-consistently into account.
Most of the expression have a transparent analytical form, which makes the
method convenient for the treatment and analysis. The separable
character of the residual interaction allows to avoid diagonalization of
high-rank RPA matrices and thus to minimize the calculation effort.
The previous studies have demonstrated high numerical accuracy and efficiency
of the method for spherical nuclei. In this contribution, the method is specified
for axial nuclei. We provide systematic and detailed presentation of
formalism and discuss different aspects of the model.}
\end{abstract}

\maketitle

\section{Introduction}
\label{sec:introduction}

Effective nuclear forces (Skyrme, Gogny, ...) are widely used
for description of diverse properties of atomic nuclei (see, for review
\cite{Ben}). However,
their application to nuclear dynamics is still limited even in the linear
regime which is usually treated within the random-phase-approximation (RPA).
The theory is is plagued by dealing with high-rank matrices which make the
computations quite expensive.  This is especially the case for non-spherical
systems with their demanding configuration space.  The rank of the matrices is
determined by the size of the one-particle-one-hole $(1ph)$ space which becomes
really huge for deformed and heavy spherical systems.

RPA problem becomes much simpler if the residual two-body interaction of a given
multipolarity $\lambda\mu$ is factorized (reduced to a separable form):
\begin{eqnarray}
\sum_{mnij}&& \!\!\! <mn|V_{res}^{\lambda\mu}|ij> a^+_m a^+_n a_j a_i \rightarrow
\sum_{k,k'=1}^{K} \kappa_{k,k'}^{\lambda\mu} {\hat X}_k^{\lambda\mu} {\hat X}_{k'}^{\lambda\mu},
\nonumber \\
&& {\hat X}_k^{\lambda\mu} = \sum_{ph} <p|{\hat X}_k^{\lambda\mu} |h> a^+_p a_h\
\label{1}
\end{eqnarray}
where ${\hat X}_k^{\lambda\mu}$ are hermitian one-body operators and
$\kappa_{k,k'}^{\lambda\mu}$ are strength constants.
The factorization  allows to reduce a high-rank RPA matrix matrix
to a small matrix with a rank determined by the number of the
separable terms. The main problem is to accomplish the factorization
self-consistently, with minimal number of separable terms and
with high accuracy.

Several self-consistent schemes \cite{BM,Row,LS,SS,Kubo,Ne_PRC,Vor,Vor,Sev}
were proposed during last decades and
signified a certain progress in this problem. However, these
schemes are not sufficiently general. Some of them are limited
to analytical or simple numerical estimates \cite{Row,BM,LS,SS},
the others are not fully self-consistent \cite{Ne_PRC}. Quite promising
is the approach \cite{Vor,Sev} for Skyrme forces. However, it still deals with
RPA matrices of rather high rank ($\sim 400$). Besides, it neglects contributions
to the residual forces from the Coulomb interaction and time-odd densities and currents .

In this connection,  we proposed  some time ago a general self-consistent
separable RPA (SRPA) approach relevant to arbitrary density- and current-dependent
functionals \cite{srpa_PRC_02,srpa_Miori,srpa_Prague,srpa_Houches}.
The method was implemented to the Skyrme functional \cite{Skyrme,Engel_75}.
In SRPA the one-body operators and strength constants of the separable expansion
are unambiguously derived from the given energy functional. Since we use the
self-consistent procedure, there is not need in any new parameters in addition to
those in the initial functional. The factorization dramatically reduces rank the of RPA
matrix. Usually a few separable terms (or even one term) are enough for accurate
reproduction of the genuine  residual interaction $V_{res}^{\lambda\mu}$ \cite{srpa_PRC_02}.
Such impressive result becomes possible due to the effective self-consistent
procedure \cite{LS} based on solid physical arguments. Besides, the separable
operators are constructed to have maxima at different slices of the nucleus and
thus cover both nuclear surface and interior \cite{srpa_PRC_02}. Hence
SRPA exhibits accuracy of most involved RPA versions but at much less expense.

One of the main SRPA advantages is its simple and transparent formalism which makes the
method very convenient for the analysis and handling of the numerical results. Being
self-consistent, SRPA allows to identify spurious admixtures connected with violation of the
translational or rotational invariance.

SRPA exploits the full 1ph space and thus equally well treats collective and non-collective
states. However, quite often (e.g. for giant resonances) we do not need
the detailed RPA description.  Then the strength function method, when we completely
avoid the calculation of the RPA states, is much more optimal. The separable
character of SRPA allows to construct the strength function based on the Lorentz
smoothing function. The strength function is naturally separated into two terms,
from the mean-field and correlations.

SRPA has a peculiarity to incorporates to the residual interaction the
contributions of both time-even and time-odd variables (densities and currents).
The time-odd variables naturally appear in the Skyrme functional constructed
to contain all the possible bilinear forms from the basic nucleon and spin densities
together with their derivatives up to the second order \cite{Dob}. Besides, time-odd
densities and currents are necessary to keep the Galilean  and gauge invariance of
the Skyrme functional \cite{Dob,Engel_75}. Skyrme functionals of this kind are
actively used for investigation of both ground state and dynamics of atomic
nuclei (see e.g. \cite{Tere} and references therein).

The recent studies with Gogny forces show that contributions of the spin-orbital
and Coulomb forces to the residual interaction can be important for the description
of low-lying states and giant resonances in exotic nuclei \cite{Peru}. SRPA
takes into account both these contributions. We plan to scrutinize their impact
in our further studies.

SRPA is quite general and in principle can be applied to a variety of
finite Fermi systems and different functionals. For example, it was derived
for the Kohn-Sham functional \cite{KS,GL} and widely used for description of linear
dynamics of valence electrons in spherical and deformed atomic clusters
\cite{Ne_PRA_98,Ne_EPJD_98,Ne_PRL_sciss,Ne_AP_02,Ne_EPJD_02,Ne_PRA_04}.

In the previous SRPA studies, we considered nuclear dynamics in spherical nuclei
\cite{srpa_PRC_02,srpa_Prague}.
However, ability of SRPA to reduce the computational effort is much more decisive for
deformed systems with its huge 1ph configuration space. So, in the present paper
we specify the Skyrme SRPA formalism to {\it axial} atomic nuclei. The pairing and its
contribution to the residual interaction are taken into account.
One of the aims of this paper is to present the SRPA by systematic and even
tutorial way, with all necessary details.

The paper is organized as follows. In Section \ref{sec:srpa}, the derivation of
the general SRPA formalism is given and discussed. The method is
SRPA is specified for Skyrme forces in Sec. \ref{sec:skyrme}.
The choice of the initial operators is discussed in Sec. \ref{sec:Q_choice}
The summary is done in Sec. \ref{sec:summary}.
Details of the formalism can be found in Appendices A-H.

\section{Basic SRPA equations}
\label{sec:srpa}

\subsection{Main requirements}

The present model provides the self-consistent factorization of the
residual interaction to the explicit form
\begin{equation} \label{V_sep}
\hat{V}_{\rm res}
\rightarrow \hat{V}_{\rm res}^{\rm sep}
= -\frac{1}{2}\sum_{ss'}\sum_{k, k'=1}^{K}
\{ \kappa_{sk,s'k'} {\hat X}_{sk} {\hat
X}_{s'k'}+\eta_{sk,s'k'} {\hat Y}_{sk} {\hat Y}_{s'k'} \} \; .
\end{equation}
We assume the residual interaction of a fixed multipolarity ($\lambda$ for spherical nuclei
and $\lambda\mu$ for deformed nuclei) but, for simplicity, skip hereafter
the multipole index. In (\ref{V_sep}), indices $s$ and $s'$ label neutrons and protons;
${\hat X}_{sk}$ and ${\hat Y}_{sk}$ are time-even
and time-odd  hermitian one-body operators.
Their time-parity properties formally read
$$
T\hat{X_{sk}}T^{-1} = \gamma_T^X \hat{X_{sk}}, \quad \gamma_T^X=+1 ,
$$
$$
T\hat{Y_{sk}}T^{-1} = \gamma_T^Y \hat{Y_{sk}}, \quad \gamma_T^Y=-1 ,
$$
where $T$ is the operator of time inversion.
The expansion (\ref{V_sep}) includes time-odd operators because some
Skyrme functionals possess both time-even
and time-odd densities and currents \cite{Ben}, see their list in the Appendix
\ref{sec:dens_curr}. As was mentioned above, the time-odd densities are necessary
to keep the Galilean  and gauge invariance of the Skyrme functional
\cite{Dob,Engel_75}. Though these densities do not contribute to the
{\it static} mean field Hamiltonian of spin-saturated systems, they can provide
time-dependent perturbations and thus have to be taken into account in the
description of nuclear dynamics.

The presence of both time-even and time-odd variables naturally leads to formulation
of the model in terms of hermitian operators with given time-parity.
These operators have the useful property that
\begin{equation}\label{eq:<[A,B]>=0}
  <0|[\hat{A},\hat{B}]|0> \sim (1-\gamma_T^A\gamma_T^B)
\end{equation}
i.e. the average commutator does not vanish only for operators
$\hat{A}$ and $\hat{B}$ having the opposite T-parities $(\gamma_T^A=-\gamma_T^B)$,
see Appendix  \ref{sec:T_par_oper} for more details. This property will be widely
used in our derivation.

The model should satisfy some principle requirements. The expansion (\ref{V_sep})
has to be self-consistent. It should involve the minimal number of the separable
terms and, at the same time, accurately reproduce the true residual interaction.
The operators ${\hat X}_k$ and ${\hat Y}_k$ and their weights should have simple
and physically transparent structure. Below we will develop the scheme which
fulfills these requirements.

\subsection{Time-dependent Hamiltonian}

The nucleus is assumed to undergo small-amplitude harmonic
vibrations around Hartree-Fock (HF) or Hartree-Fock-Bogoliubov
(HFB) ground state. We start with a
general time-dependent  functional $E(J_{s}^{\alpha}({\vec
r},t))$ involving a set of arbitrary neutron and proton
densities and currents $J_{s}^{\alpha}({\vec r},t)$ (where $s=n,p$ ;
$\alpha$ labels densities and currents)
\begin{equation}\label{eq:E(J)}
E(J_{s}^{\alpha}({\vec r},t))=
<\Psi(t) |\hat H| \Psi(t)>=\int {\cal H} ({\vec r})d{\vec r}
\label{2}
\end{equation}
where $|\Psi(t)\!>$ is the wave function of the vibrating system
described as the time-dependent Slater determinant.
Time-dependent densities and currents are determined through
the corresponding operators as
\begin{equation}\label{eq:dens_J}
J_{s}^{\alpha}({\vec r},t)=
_s<\Psi(t) |\hat J_{s}^{\alpha}({\vec r})| \Psi(t)>_s
= \sum_{h \epsilon s}^{occ}
 \varphi^*_h({\vec r},t)\hat J_{\alpha}({\vec r})\varphi_h^{\mbox{}}({\vec r},t)
\label{J}
\end{equation}
where $\varphi_h^{\mbox{}}({\vec r},t)$ is wave function of the
hole (occupied) single-particle state. The set (\ref{J}) includes
both time-even  and time-odd
densities and currents, see examples in the Appendix \ref{sec:dens_curr}.

In the linear regime, the time-dependent density reads as a sum of the static
part and the small time-dependent perturbation:
\begin{equation}\label{eq:dens_J(t)}
J_{s}^{\alpha}({\vec r},t) =  \bar{J}_{s}^{\alpha}({\vec r})+
\delta J_{s}^{\alpha}({\vec r},t)
\end{equation}

Then, substituting (\ref{eq:dens_J(t)}) into (\ref{eq:E(J)}) and keeping the terms
up to the linear order for $\delta J_{s}^{\alpha}({\vec r},t)$,
one gets the single-particle Hamiltonian
\begin{equation}\label{full_h}
 \hat h (t)  = \hat{h}_0 + \hat{h}_{\rm res}(t)
\end{equation}
with the static mean-field part
\begin{eqnarray}\label{eq:h_0}
\hat{h}_0  = \sum_{\alpha,s} \frac{\delta E
(J_{n}^{\alpha},J_{p}^{\alpha})} {\delta J_{s}^{\alpha}}
\hat{J}_{s}^{\alpha}
\end{eqnarray}
and the time-dependent response
\begin{equation}\label{eq:h_resp}
\hat{h}_{\rm res}(t)  = \sum_{\alpha's'}
[\frac{\delta {\hat h}_0}{\delta J_{s'}^{\alpha'}}]_{J=\bar{J}}
\delta J_{s'}^{\alpha'}(t)
= \sum_{\alpha s}\sum_{\alpha's'}
[\frac{\delta^2 E(\vec r)}
{\delta J_{s}^{\alpha}\delta J_{s'}^{\alpha'}}]_{J=\bar{J}}
\delta J_{s'}^{\alpha'}(t)
{\hat J}_{s}^{\alpha} \; .
\end{equation}
The later determines oscillations of the system. For the brevity of notation, we skip
the dependence on space coordinates
in (\ref{full_h})-(\ref{eq:h_resp}) and hereafter in this section.
The explicit space dependence of the key
SRPA operators and values can be found in the Sec. \ref{sec:skyrme}.

The next step in our derivation is to specify the unknown density variations
(or transition densities)
\begin{equation}\label{eq:J(t)_psi}
\delta J_{s}^{\alpha}(t)  =
<\Psi(t)|{\hat J}_{s}^{\alpha}|\Psi(t)> -
<0|{\hat J}_{s}^{\alpha}|0>
\end{equation}
where $|0>$ is the static ground state. For this aim we should define
the perturbed many-body wave function $|\Psi(t)>$.

\subsection{Scaling perturbed wave function}

The macroscopic perturbed many-body wave function $|\Psi(t)\!>_s$ is obtained
from the static HF or HFB ground state $|0 >_s$ by the scaling
transformation \cite{LS}
\begin{equation}\label{eq:scaling}
|\Psi(t)\!>_s=\prod_{k=1}^K
exp [-iq_{sk}(t)\hat{P}_{s k}]
exp [-ip_{sk}(t)\hat{Q}_{s k}]
|0>_s .
\end{equation}
Here both $|\Psi(t)\! >_s $ $|0 >_s$ are Slater determinants;
${\hat Q}_{sk}(\vec{r})$ and
${\hat P}_{sk}(\vec{r})$ are generalized coordinate (time-even) and
momentum (time-odd) hermitian operators with the properties
\begin{eqnarray}\label{eq:P_Q}
\hat{Q}_{sk} =\hat{Q}_{sk}^+,\quad \gamma_T^Q=1, \nonumber \\
\hat{P}_{sk} = i[\hat{H},\hat{Q}_{sk}]_{ph}=\hat{P}_{sk}^+,\quad
\gamma_T^P=-1
\end{eqnarray}
where $\hat{H}=\hat{h}_0 +\hat{V}_{\rm res}$ is the full Hamiltonian
embracing both one-body and two-body parts. The subscript $ph$ in
the commutator means the mapping into particle-hole domain.
If the functional includes only time-even densities, then
$\hat{V}_{\rm res}$ does not contribute to the commutator and so one
may use $\hat{h}_0$ instead of $\hat{H}$.

Operators (\ref{eq:P_Q})
generate time-even and time-odd real
collective deformations $q_{sk}(t)$ and $p_{sk}(t)$.
Using (\ref{eq:scaling}), the transition densities (\ref{eq:J(t)_psi}) are
expressed through these deformations as
\begin{eqnarray}\label{eq:trans_dens}
\delta J_{s}^{\alpha}(t)&&  =
<\Psi(t)|{\hat J}_{s}^{\alpha}|\Psi(t)> -
<0|{\hat J}_{s}^{\alpha}|0>  =  \\
&&  =
i \sum_k
\{ q_{sk}(t)<0|[\hat{P}_{s k},\hat J_{s}^{\alpha}]|0>
+ p_{s k}(t)<0|[\hat{Q}_{s k},\hat J_{s}^{\alpha}]|0>\} \; .
\nonumber
\end{eqnarray}
Then the response Hamiltonian (\ref{eq:h_resp}) can be recast as
\begin{equation}\label{eq:h_XY}
\hat{h}_{\rm res}(t)  = \sum_{s k}
\{ q_{sk}(t) \hat{X}_{s k} +
 p_{s k}(t) \hat{Y}_{s k}\}
\end{equation}
where all time-independent terms are collected
in the hermitian one-body operators
\begin{eqnarray}\label{eq:X}
\hat{X}_{s k}  = \sum_{s'}\hat{X}_{s k}^{s'} =
\sum_{s'} i\sum_{\alpha' \alpha}
[\frac{\delta^2 E} {\delta J_{s'}^{\alpha'}
\delta J_{s}^{\alpha}}]_{J=\bar{J}}
<0|[\hat{P}_{s k} ,{\hat J}_{s}^{\alpha}]|0>
{\hat J}_{s'}^{\alpha'} ,
\\
\hat{Y}_{s k}   = \sum_{s'}\hat{Y}_{s k}^{s'} =
\sum_{s'} i\sum_{\alpha' \alpha}
[\frac{\delta^2 E} {\delta J_{s'}^{\alpha'}
\delta J_{s}^{\alpha}}]_{J=\bar{J}}
<0|[\hat{Q}_{s k} ,{\hat J}_{s}^{\alpha}]|0>
{\hat J}_{s'}^{\alpha'}
\label{eq:Y}
\end{eqnarray}
with the properties
\begin{eqnarray}\label{X-properties}
\hat{X}_{sk} &=& \hat{X}_{sk}^+,  \quad
\gamma_T^X=+1, \quad
\hat{X}^* = \hat{X},
\\
\label{Y-properties}
\hat{Y}_{sk} &=& \hat{Y}_{sk}^+, \quad
\gamma_T^Y=-1, \quad
\hat{Y}^*=-\hat{Y} .
\end{eqnarray}
As is shown below, ${\hat X}_{sk}$  and ${\hat Y}_{sk}$ are just the time-even and
time-odd operators to be exploited in the separable expansion (\ref{V_sep}).
Following the  property (\ref{eq:<[A,B]>=0}), time-even densities contribute only
to ${\hat X}_{sk}$ while time-odd densities only to ${\hat Y}_{sk}$.
The upper index $s'$ in the operators (\ref{eq:X})-(\ref{eq:Y}) determines the
isospin (proton or neutron) subspace where these operators act.
This is the domain of the density operator ${\hat J}_{s'}^{\alpha'}$
entering (\ref{eq:X})-(\ref{eq:Y}).

To complete the construction of the separable expansion (\ref{V_sep}),
we should still determine the matrices of the strength constants
$\kappa_{sk,s'k'}$ and  $\eta_{sk,s'k'}$. This can be done through
variations of the basic operators
(\ref{eq:trans_dens}):
\begin{eqnarray}\label{eq:Xvar}
\delta {\hat X}_{s k}(t)  && \equiv
<\Psi(t)|{\hat X}_{sk}|\Psi(t)>-<0|{\hat X}_{sk}|0> =
\\
&& = i \sum_{s'k'}
q_{s'k'}(t)<0|[\hat{P}_{s'k'},\hat X_{sk}^{s'}]|0> =
- \sum_{s'k'} q_{s'k'}(t) \kappa_{s'k',sk}^{-1} \; ,
\nonumber
\end{eqnarray}
\begin{eqnarray}\label{eq:Yvar}
\delta {\hat Y}_{s k}(t)  && \equiv
<\Psi(t)|{\hat Y}_{sk}|\Psi(t)>-<0|{\hat Y}_{sk}|0> =
\\
&& = i \sum_{s'k'} p_{s'k'}(t) <0|[\hat{Q}_{s'k'},\hat Y_{sk}^{s'}]|0> =
- \sum_{s'k'} p_{s'k'}(t) \eta_{s'k',sk}^{-1}
\nonumber
\end{eqnarray}
where
\begin{eqnarray}
\label{eq:kappa}
\kappa_{s'k',sk}^{-1 }&&=
\kappa_{sk,s'k'}^{-1} =
- i <0|[\hat{P}_{s'k'},{\hat X}_{sk}^{s'}]|0> = \\
&& = \int d{\vec r}\sum_{\alpha \alpha'}
[\frac{\delta^2 E}{\delta J_{s'}^{\alpha'}\delta J_{s}^{\alpha}}]
<0|[\hat{P}_{s k},{\hat J}_{s}^{\alpha}]|0>
<0|[\hat{P}_{s' k'},{\hat J}_{s'}^{\alpha'}]|0> \; ,
\nonumber
\end{eqnarray}
\begin{eqnarray}
\label{eq:eta}
\eta_{s'k',sk}^{-1 }&&=
\eta_{sk,s'k'}^{-1 } = -i
<0|[\hat{Q}_{s'k'},{\hat Y}_{sk}^{s'}]|0>  \\
&& = \int d{\vec r}\sum_{\alpha \alpha'}
[\frac{\delta^2 E} {\delta J_{s'}^{\alpha'}\delta J_{s}^{\alpha}}]
<0|[\hat{Q}_{s k},{\hat J}_{s}^{\alpha}]|0>
<0|[\hat{Q}_{s' k'},{\hat J}_{s'}^{\alpha'}]|0>.  \nonumber
\end{eqnarray}
Eqs. (\ref{eq:kappa})-(\ref{eq:eta})
represent elements of the symmetric matrix which is inverse
to the matrix of the strength constants in (\ref{V_sep}).
Indeed Eqs. (\ref{eq:Xvar})-(\ref{eq:Yvar}) can be recast to
\begin{eqnarray}
- \sum_{sk} \kappa_{s'k',sk} \delta {\hat X}_{s k}(t) &=& q_{s'k'}(t) \; ,
\\
-  \sum_{sk} \eta_{s'k',sk} \delta {\hat Y}_{s k}(t) &=& p_{s'k'}(t) \; .
\end{eqnarray}
Then the response Hamiltonian (\ref{eq:h_XY})
gains the form
\begin{equation}\label{eq:h_dXdY}
\hat{h}_{\rm res}(t)  =  - \sum_{s'k'} \sum_{sk}
\{
\kappa_{s'k',sk} \delta\hat{X}_{sk}(t) \hat{X}_{s'k'}+
\eta_{s'k',sk} \delta\hat{Y}_{sk}(t) \hat{Y}_{s'k'}
\}
\end{equation}
which leads to the same eigenvalue problem  as the separable Hamiltonian
\begin{equation}
{\hat H}={\hat h}_0 + {\hat V}^{\rm sep}_{\rm res}
\label{eq:H_sep}
\end{equation}
with ${\hat V}^{\rm sep}_{\rm res}$ from (\ref{V_sep}).
See also \cite{Row}) for relevant discussion.

In principle, we have  already in our disposal the formalism for
linear regime of the collective motion in terms of collective
harmonic variables
\begin{eqnarray}
q_{sk}(t)=\bar{q}_{s k} \cos(\omega t)
=\frac{1}{2}\bar{q}_{s k}(e^{i\omega t}+e^{-i\omega t}) \; ,
\\
p_{sk}(t)=\bar{p}_{s k} sin(\omega t)=
\frac{1}{2i}\bar{p}_{s k}(e^{i\omega t}-e^{-i\omega t}) \; .
\label{eq:qp_harm}
\end{eqnarray}
Indeed, Eqs. (\ref{eq:X}), (\ref{eq:Y}), (\ref{eq:kappa}), and (\ref{eq:eta})
deliver the one-body operators and strength matrices which we need for the
separable expansion of the two-body interaction. The substitution of
the response Hamiltonian (\ref{eq:h_dXdY}) and the perturbed wave function
(\ref{eq:scaling}) into time-dependent HF equation
\begin{equation}
i\frac{d}{dt}|\Psi(t)>=
({\hat h}_0+{\hat h}_{\rm res}(t))|\Psi(t)>
\label{eq:HF}
\end{equation}
would result in the eigenvalue problem. The number $K$ of the
collective variables (and thus of the separable terms)
depends on the accuracy we need in the description of collective modes
(see discussion in Sec. \ref{sec:Q_choice}).
For $K=1$, the method reduces to the sum rule approach with one collective mode
\cite{Re_AP_92}.  For $K>1$, we have a system of K coupled oscillators and
the method is reduced to so-called local RPA \cite{Re_PRA_90,Re_AP_92} suitable for a
rough description of main branching and gross-structure properties of collective modes.
However, the method is still not ready to describe the Landau fragmentation. For this
aim, we should consider the detailed 1ph space. This will be done in the next
subsection.

\subsection{Coupling with $1ph$ space}

Collective modes can be viewed as superpositions of 1ph configurations.
To derive this relation, it is convenient to introduce an
alternative perturbed many-body wave function
\begin{equation}
|\Psi(t)> = (1+\sum_s \sum_{ph\epsilon s} c^s_{ph}(t) \hat{A}^+_{ph}) |0>
\label{Psi_Thouless}
\end{equation}
where
\begin{equation}\label{eq:1ph}
 \hat{A}^+_{ph}=a^{\dagger}_p a_h
\end{equation}
is operator of the creation of 1ph pair and
\begin{equation}
c^s_{ph}(t)=c^{s+}_{ph}e^{i\omega t}+c^{s-}_{ph}e^{-i\omega t}
\label{eq:c_ph}
\end{equation}
are time-dependent amplitudes of particle-hole configurations
in the perturbed state. Here we used the Thouless theorem \cite{Thouless} which
establishes the connection between two arbitrary Slater determinants.
The wave function (\ref{Psi_Thouless})
is obviously the {\it microscopic} counterpart of the {\it macroscopic} wave function
(\ref{eq:scaling}).

Substituting  (\ref{eq:h_XY})  and (\ref{Psi_Thouless})
into the time-dependent HF equation (\ref{eq:HF}), one gets,
in the linear approximation,
the relation between $c^{\pm}_{ph}$ and collective deformations
$\bar{q}_{sk}$ and $\bar{p}_{sk}$
\begin{equation}
c^{s\pm}_{ph \epsilon s}=-\frac{1}{2}
\frac{\sum_{s'k'} [\bar{q}_{s'k'}
<ph|\hat{X}^s_{s'k'}|0>
\mp i \bar{p}_{s'k'}<ph|\hat{Y}^s_{s'k'}|0>]}
{\varepsilon_{ph}\pm\omega},
\label{eq:c_pm_qp}
\end{equation}
where $\varepsilon_{ph}$ is the energy of 1ph pair.

In addition to Eqs. (\ref{eq:Xvar})-(\ref{eq:Yvar}), the variations
$\delta {\hat X}_{sk}(t)$  and $\delta {\hat Y}_{sk}(t)$
can be now  obtained with the microscopic perturbed wave function
(\ref{Psi_Thouless}):
\begin{eqnarray}\label{eq:dX_ph}
\delta {\hat X}_{sk}(t) &=&
\sum_{s'} \sum_{ph \epsilon s'}
(c^{s'}_{ph}(t)^*<\!ph|\hat{X}^{s'}_{sk}|0\!>
+ c^{s'}_{ph}(t)<\!0|\hat{X}^{s'}_{sk}|ph\!>),
 \\
\delta {\hat Y}_{sk}(t) &= &
\sum_{s'} \sum_{ph \epsilon s'}
(c^{s'}_{ph}(t)^*<\!ph|\hat{Y}^{s'}_{sk}|0\!>
+ c^{s'}_{ph}(t)<\!0|\hat{Y}^{s'}_{sk}|ph\!>) \; .
\label{eq:dY_ph}
\end{eqnarray}

\subsection{Eigenvalue problem}

But both amplitudes $c^{s\pm}_{ph}$
and collective variables $\bar{q}_{sk}$ and $\bar{p}_{sk}$ are still unknown.
The time-dependent HF equation was already exploited and cannot be used
once more to determine these unknowns. Thus we need for this aim
some additional physical constraint. It can be naturally formulated
as equality of the dynamical variations of the basic operators
$\delta{\hat X}_{k}$ and $\delta{\hat Y}_{k}$, obtained with the macroscopic
(\ref{eq:scaling}) and microscopic (\ref{Psi_Thouless})  perturbed wave functions.
Thus we should equate (\ref{eq:Xvar})-(\ref{eq:Yvar})
and (\ref{eq:dX_ph})-(\ref{eq:dY_ph}). This gives
\begin{eqnarray}
\label{eq:X_c_qp}
- \sum_{s'k'} q_{s'k'}(t) \kappa_{s'k',sk}^{-1}
&=&
\sum_{s'} \sum_{ph \epsilon s'}
(c^{s'}_{ph}(t)^*<\!ph|\hat{X}^{s'}_{sk}|0\!>
+c^{s'}_{ph}(t)<\!0|\hat{X}^{s'}_{sk}|ph\!>),
\\
- \sum_{s'k'} q_{s'k'}(t) \eta_{s'k',sk}^{-1}
&=&
 \sum_{s'} \sum_{ph \epsilon s'}
(c^{s'}_{ph}(t)^*<\!ph|\hat{Y}^{s'}_{sk}|0\!>+
c^{s'}_{ph}(t)<\!0|\hat{Y}^{s'}_{sk}|ph\!>).
\label{eq:Y_c_qp}
\end{eqnarray}
Substituting  (\ref{eq:qp_harm}) and  (\ref{eq:c_ph})
into these expressions and collecting, for example, the terms with
$e^{i\omega t}$, we gets
\begin{eqnarray}
\label{RPA_system_eq}
- \sum_{s'k'} \bar{q}_{s'k'} \kappa_{s'k',sk}^{-1}
&=&
2 \sum_{s'} \sum_{ph \epsilon s'}
\{
(c^{s' -}_{ph})^*<\!ph|\hat{X}^{s'}_{sk}|0\!>
+ c^{s' +}_{ph}<\!0|\hat{X}^{s'}_{sk}|ph\!>
\} ,
\\
i \sum_{s'k'} \bar{p}_{s'k'} \eta_{s'k',sk}^{-1}
&=&
 2 \sum_{s'} \sum_{ph \epsilon s'}
\{
(c^{s' -}_{ph})^*<\!ph|\hat{Y}^{s'}_{sk}|0\!>
+ c^{s' +}_{ph}<\!0|\hat{Y}^{s'}_{sk}|ph\!>
\} .
\label{eq:Y_c_qp2}
\end{eqnarray}
Then, by using (\ref{eq:c_pm_qp}), all the unknowns
in the expressions are reduced  to the collective variables
$\bar{q}_{k}$ and  $\bar{p}_{k}$ and we  finally get
 \begin{eqnarray}
\label{eq:X_c_qp4}
\sum_{\bar{s}\bar{k}}  &
\{
\bar{q}_{\bar{s}\bar{k}}
[F_{s'k',\bar{s}\bar{k}}^{(XX)}- \kappa_{\bar{s}\bar{k},s'k'}^{-1}]
+\bar{p}_{\bar{s}\bar{k}} F_{s'k',\bar{s}\bar{k}}^{(XY)}
\}=0 \; ,
\nonumber\\
\sum_{\bar{s}\bar{k}}  &
\{
\bar{q}_{\bar{s}\bar{k}}
F_{s'k',\bar{s}\bar{k}}^{(YX)}
+\bar{p}_{\bar{s}\bar{k}}
[F_{s'k',\bar{s}\bar{k}}^{(YY)} - \eta_{\bar{s}\bar{k},s'k'}^{-1}]
\}=0
\end{eqnarray}
with
\begin{eqnarray}
F_{s'k',\bar{s}\bar{k}}^{(XX)} && =
 \sum_s \sum_{ph \epsilon s}
\frac{1}{\varepsilon_{ph}^2-\omega^2}
\{<ph|\hat{X}^s_{\bar{s}\bar{k}}|0>^* <ph|\hat{X}^s_{s'k'}|0>
(\varepsilon_{ph}+\omega) \nonumber \\
&& \hspace{3.5cm}+
<ph|\hat{X}^s_{\bar{s}\bar{k}}|0> <0|\hat{X}^s_{s'k'}|ph>
(\varepsilon_{ph}-\omega)\} ,  \\
F_{s'k',\bar{s}\bar{k}}^{(YX)} && =
-i \sum_s \sum_{ph \epsilon s}
\frac{1}{\varepsilon_{ph}^2-\omega^2}
\{<ph|\hat{Y}^s_{\bar{s}\bar{k}}|0>^* <ph|\hat{X}^s_{s'k'}|0>
(\varepsilon_{ph}+\omega)  \nonumber\\
&& \hspace{3.5cm}+
<ph|\hat{Y}^s_{\bar{s}\bar{k}}|0> <0|\hat{X}^s_{s'k'}|ph>
(\varepsilon_{ph}-\omega)\} , \\
F_{s'k',\bar{s}\bar{k}}^{(XY)} && =
i \sum_s \sum_{ph \epsilon s}
\frac{1}{\varepsilon_{ph}^2-\omega^2}
 \{<ph|\hat{X}^s_{\bar{s}\bar{k}}|0>^* <ph|\hat{Y}^s_{s'k'}|0>
(\varepsilon_{ph}+\omega) \nonumber \\
&& \hspace{3.5cm}  +
<ph|\hat{X}^s_{\bar{s}\bar{k}}|h> <0|\hat{Y}^s_{s'k'}|ph>
(\varepsilon_{ph}-\omega)\} ,
\\
F_{s'k',\bar{s}\bar{k}}^{(YY)} && =
  \sum_s \sum_{ph \epsilon s}
\frac{1}{\varepsilon_{ph}^2-\omega^2}
\{<ph|\hat{Y}^s_{\bar{s}\bar{k}}|0>^* <ph|\hat{Y}^s_{s'k'}|0>
(\varepsilon_{ph}+\omega) \nonumber \\
&& \hspace{3.5cm}+
<ph|\hat{Y}^s_{\bar{s}\bar{k}}|0> <0|\hat{Y}^s_{s'k'}|ph>
(\varepsilon_{ph}-\omega)\} .
\label{eq:F_1}
\end{eqnarray}
These equations can be simplified using the relations
\begin{eqnarray}
   <0|\hat{X}^s_{\bar{s}\bar{k}}|ph>
= <ph|\hat{X}^s_{\bar{s}\bar{k}}|0>^*
= <ph|\hat{X}^s_{\bar{s}\bar{k}}|0>
= \overline{<ph|\hat{X}^s_{\bar{s}\bar{k}}|0>}
,
\\
   <0|\hat{Y}^s_{\bar{s}\bar{k}}|ph>
= <ph|\hat{Y}^s_{\bar{s}\bar{k}}|0>^*
= - <ph|\hat{Y}^s_{\bar{s}\bar{k}}|0>
= - i \overline{<ph|\hat{Y}^s_{\bar{s}\bar{k}}|0>}
\end{eqnarray}
which directly follow from the properties (\ref{X-properties}) and
(\ref{Y-properties}) and expressions (\ref{eq:X}) and  (\ref{eq:Y}).
Then elements of the RPA matrix can be rewritten in terms of
real (overline) matrix elements as
\begin{eqnarray}\label{eq:F_XX}
F_{s'k',\bar{s}\bar{k}}^{(XX)} && =
 2 \sum_s \sum_{ph \epsilon s}
\frac{\varepsilon_{ph}
\overline{<ph|\hat{X}^s_{\bar{s}\bar{k}}|0>}
\; \overline{<ph|\hat{X}^s_{s'k'}|0>}}
{\varepsilon_{ph}^2-\omega^2},
\\
\label{eq:F_YX}
F_{s'k',\bar{s}\bar{k}}^{(YX)} = F_{s'k',\bar{s}\bar{k}}^{(XY)} && =
-2 \sum_s \sum_{ph \epsilon s}
\frac{\omega
\overline{<ph|\hat{Y}^s_{\bar{s}\bar{k}}|0>}
\; \overline{<ph|\hat{X}^s_{s'k'}|0>}}
{\varepsilon_{ph}^2-\omega^2} ,
%
\\
\label{eq:F_YY}
F_{s'k',\bar{s}\bar{k}}^{(YY)} && =
 2 \sum_s \sum_{ph \epsilon s}
\frac{\varepsilon_{ph}
\overline{<ph|\hat{Y}^s_{\bar{s}\bar{k}}|0>}
\; \overline{<ph|\hat{Y}^s_{s'k'}|0>}}
{\varepsilon_{ph}^2-\omega^2} .
\end{eqnarray}

Supposing determinant of the system (\ref{eq:X_c_qp4}) to be zero,
we obtain the dispersion equation for determination of the RPA
eigenvalues $\omega_{\nu}$.

\subsection{Normalization condition}

The standard RPA operators of excited one-phonon states $|\nu>$ are
defined as
\begin{equation}\label{eq:Q_op}
\hat{Q}^+_{\nu}= \frac{1}{2}\sum_s \sum_{ph \epsilon s}
\{
c^{\nu s -}_{ph} \hat{A}^+_{ph} - c^{\nu s +}_{ph} \hat{A}_{ph}
\}
\end{equation}
and fulfill the normalization and orthogonality conditions
\begin{equation}\label{eq:Q_norm}
\langle[\hat{Q}_{\nu},\hat{Q}_{\nu'}^+]\rangle = \delta_{\nu,\nu'}
, \quad
\langle[\hat{Q}_{\nu}^+,\hat{Q}_{\nu'}^+]\rangle
= \langle[\hat{Q}_{\nu},\hat{Q}_{\nu'}]\rangle = 0 ,
\end{equation}
where $\hat{A}^+_{ph}$ and $c^{\nu \pm}_{ph}$ are given by (\ref{eq:1ph})
and (\ref{eq:c_pm_qp}), respectively.  In the quasi-boson approximation
($[\hat{A}_{ph},\,\hat{A}^+_{p'h'}]=\delta_{pp'}\delta_{hh'}$),
the normalization condition
$[\hat{Q}_{\nu},\hat{Q}_{\nu}^+] = 1$ results in the relation
\begin{equation}\label{eq:c_norm}
\sum_s \sum_{ph \epsilon s}
\{
(c^{\nu s-}_{ph})^2 - (c^{\nu s+}_{ph})^2
\} =2 .
\end{equation}
By using (\ref{eq:c_pm_qp}),  it can be recast in terms
of the RPA matrix coefficients (\ref{eq:F_XX})-(\ref{eq:F_YY}):
\begin{eqnarray}\label{eq:c_norm_F}
&& \sum_s \sum_{ph \epsilon s}
\{
(c^{\nu s-}_{ph})^2 - (c^{\nu s+}_{ph})^2
\}
 \\
&& =  \sum_{s'k'}  \sum_{\bar{s}\bar{k}}\frac{1}{4}
 \{
\bar{q}^{\nu}_{s'k'} \bar{q}^{\nu}_{\bar{s}\bar{k}}
\frac{\partial F_{s'k',\bar{s}\bar{k}}^{(XX)}(\omega_{\nu})}{\partial\omega_{\nu}}
+
2 \bar{q}^{\nu}_{s'k'} \bar{p}^{\nu}_{\bar{s}\bar{k}}
\frac{\partial F_{s'k',\bar{s}\bar{k}}^{(YX)}(\omega_{\nu})}{\partial\omega_{\nu}}
+
\bar{p}^{\nu}_{s'k'} \bar{p}^{\nu}_{\bar{s}\bar{k}}
\frac{\partial F_{s'k',\bar{s}\bar{k}}^{(YY)}(\omega_{\nu})}{\partial\omega_{\nu}}
\} = 2 N_{\nu} .
\nonumber
\end{eqnarray}
Hence the variables  $\bar{q}^{\nu}_{sk}$ and
$\bar{p}^{\nu}_{sk}$ should be renormalized
by the factor $1/\sqrt{N_{\nu}}$.

\subsection{Pairing contribution}
\label{sec:pairing}

The pairing Hamiltonian reads
\begin{equation}\label{pair_hamilt}
\hat h_{\rm pair} \,=\, - \sum_{s=n,p} G_s \hat{\chi}_s^{+} \hat{\chi}_s
\end{equation}
where
\begin{equation}\label{chi}
\hat{\chi}_s^{+} = \sum_{j \epsilon s} \, a_j^+ a_{\bar j}^+
\qquad
\hat{\chi}_s = \sum_{j \epsilon s} \, a_{\bar j} a_j .
\end{equation}
Then, after the Bogoliubov transformation from particle to quasiparticle operators,
the hermitian one-body operators have a general form
\begin{eqnarray}
\hat A &=& \sum_{ij} <ij|A|0> ( a^+_{i} a_{j} +  a^+_{j} a_{i})
\\
&=& 2 \sum_{ij}^{K_j,K_i>0}
\{
<ij|A|0>
(u_i v_j + \gamma^A_T u_j v_i)
    (\hat{\alpha}^+_{i} \hat{\alpha}^+_{\bar j}
   + \gamma^A_T \hat{\alpha}_{\bar i}\hat{\alpha}_{j})
\nonumber \\
&& \qquad\qquad \qquad\qquad \qquad
+(u_i u_j - \gamma^A_T v_j v_i)
(\hat{\alpha}^+_{j} \hat{\alpha}_{i}
 - \gamma^A_T
    \hat{\alpha}_{\bar i} \hat{\alpha}^+_{\bar j})
\}
\label{eq:A_2qp_ph_Ki-Kj}
\end{eqnarray}
 where $\bar i$ are time-reversed states and the time-inverse factor $\gamma_T^A$
 defines time-parity of the operator $\hat A$ (see details in
 Appendix \ref{sec:herm+time-conj}).

Hence the time-even $(\gamma_T^A=1)$ and time-odd
$(\gamma_T^B=-1)$ operators read
\begin{eqnarray}\label{eq:Q_2qp}
{\hat A} &=&  2 \sum_{ij}^{K_j,K_i>0}
<ij|A|0>
(u_i v_j + u_j v_i)
    (\hat{\alpha}^+_{i} \hat{\alpha}^+_{\bar j}
   +  \hat{\alpha}_{\bar i}\hat{\alpha}_{j}) \; ,
\\
{\hat B} &=&  2 \sum_{ij}^{K_j,K_i>0}
<ij|B|0> (u_i v_j - u_j v_i)
    (\hat{\alpha}^+_{i} \hat{\alpha}^+_{\bar j}
   -  \hat{\alpha}_{\bar i}\hat{\alpha}_{j}) \; ,
\label{eq:P_2qp}
\end{eqnarray}
i.e. obtain the pairing factors
\begin{equation}\label{pairing_factors}
u^{(+)}_{ij}=u_i v_j + u_j v_i , \quad u^{(-)}_{ij}=u_i v_j - u_j v_i .
\end{equation}
This is the case of time even-operators $\hat Q_{sk}$ and $\hat X_{sk}$ and
the time-odd operator $\hat Y_{sk}$. The situation with the time-odd operator
$$
\hat P_{sk}=i[\hat H, \hat Q_{sk}]
=i \{ [\hat h_{0}, \hat Q_{sk}] + [\hat V_{\rm res}^{\rm sep}, \hat Q_{sk}] \}
= i [\hat h_{0}, \hat Q_{sk}] - \hat Y_{sk}
$$
is more complicated because of the additional term $i[\hat h_{0}, \hat Q_{sk}]$.
Taking into account
(\ref{eq:Q_2qp}) and (\ref{eq:h_HFB})
this operator reads
\begin{equation}\label{eq:op_P_comm}
\hat{P}_{sk}
=
2 \sum_{ij \epsilon s}^{K_j,K_i>0}
\{
i2 \epsilon_{ij}  u^{(+)}_{ij} <ij|Q_{sk}|0>
-  u^{(-)}_{ij}  <ij|Y_{sk}^s|0>
\} \; (\hat{\alpha}^+_{i} \hat{\alpha}^+_{\bar j}
   -  \hat{\alpha}_{\bar i}\hat{\alpha}_{j}) .
\end{equation}
It is seen that $\hat{P}_{sk}$ keeps the same operator structure
$\hat{\alpha}^+_{i} \hat{\alpha}^+_{\bar j} -  \hat{\alpha}_{\bar i}\hat{\alpha}_{j}$
as (\ref{eq:P_2qp}) but, at the same has, unlike $\hat{Y}_{sk}$, the diagonal
($i=j$ matrix elements. The later is the obvious consequence of the term
$i[\hat h_{0}, \hat Q_{sk}]$. Explicit expressions for the matrix elements of
the operator $\hat{P}_{sk}$ are given in the Appendix \ref{sec:matr_el_P}.

Besides the factors (\ref{pairing_factors}) the pairing results in the specific
contribution to the time-even part of the response Hamiltonian $\hat h_{\rm res}(t)$.
This contribution can  be derived in close analogy with other time-even densities.
Then we get
\begin{equation} \label{h_quadr_pair}
\delta \hat h_{\rm res}^{\rm pair}(t) \, = \sum_{ks} q_{ks}(t)\sum_{s} \hat X^{s(pair)}_{ks}
\nonumber
\end{equation}
where
\begin{equation}
\hat X_{ks}^{s(pair)} = - G_s \chi_{X, sk}\sum_{j \epsilon s}(u_j^2 - v_j^2)
(\alpha_{\bar j} \alpha_j \,+\, \alpha_{\bar j}^{+} \alpha_j^{+}) .
\label{X_pair}
\end{equation}
The pairing response reads
\begin{eqnarray}
\chi_{X, sk} &=& i \, <\tilde 0| [{\hat P}_{ks}, {\hat \chi}_s^+ + {\hat \chi}_s ]|\tilde 0>
\nonumber \\
&=& i \,
\sum_{j\varepsilon s}
8i \epsilon_j u^{(+)}_{jj} <j\bar j|Q_{ks}|0>
\sum_{l\varepsilon s}(u_l^2 - v_l^2)
<\tilde 0| [ (\alpha_j^{+} \alpha_{\bar j}^{+} - \alpha_{\bar j} \alpha_j),
(\alpha_{\bar l} \alpha_l \,+\, \alpha_{\bar l}^{+} \alpha_l^{+}) ]|\tilde 0>
\nonumber \\
&=&
16 \sum_{j\varepsilon s}
 \varepsilon_{j} \,(u_j^2 - v_j^2 ) \,
u^{(+)}_{jj}  <j\bar j|Q_{ks}|0>
\label{pair_resp}
\end{eqnarray}
where we used the diagonal part of the operator (\ref{eq:op_P_comm}).
Finally, the pairing contribution to the strength matrix is
\begin{equation}\label{kappa_pair}
[\kappa^{(pair)}_{sk'sk}]^{-1} =
-i<\tilde 0|[{\hat P}_{k's}, {\hat X}_{sk}^{s (pair)} ]|\tilde 0>
= G_s \chi_{X, sk'} \chi_{X, sk} .
\end{equation}

It worth noting that in all the sections except of the present one, the pairing
factors (\ref{pairing_factors}) are supposed to be included into the matrix elements
and transition densities and thus are not depicted explicitly.

\subsection{Strength function method}
\label{sec:strength_func}

In exploration of the system response to external fields, we are usually
interested in the total strength function instead of the responses
of particular RPA states. For example, giant resonances in heavy nuclei
are formed by thousands of RPA states whose contributions in any case cannot
be distinguished experimentally. In this case, it is reasonable
to implement the strength function formalism.
Besides, the calculation of the strength function is much easier.

For electric external fields of multipolarity $E\lambda\mu$, the strength
function can be defined as
\begin{equation}\label{eq:strength_function}
S_L(E\lambda\mu ; \omega)= \sum_{\nu}
\omega_{\nu}^{L} M_{\lambda\mu \nu}^2 \zeta(\omega - \omega_{\nu})
\end{equation}
where
\begin{equation}
\zeta(\omega - \omega_{\nu}) = \frac{1}{2\pi}
  \frac{\Delta}{(\omega - \omega_{\nu})^2 + (\Delta/2)^2}              
\label{eq:lorfold}
\end{equation}
is Lorentz weight with an averaging parameter $\Delta$
and
\begin{equation}\label{eq:tr_me}
M_{\lambda\mu \nu} = \frac{1}{\sqrt{2}}\sum_s e^{eff}_s
\sum_{ij \epsilon {s}}
<ij|\hat f_{\lambda\mu}|0> ( c^{\nu s-}_{ij} + c^{\nu s+}_{ij})
\end{equation}
is the matrix element of $E\lambda\mu$ transition from the ground state
to the RPA state $|\nu>$. The pairing factor is included to the matrix elements.
The forward and backward two-quasiparticle
amplitudes $c^{\nu s\pm}_{ij}$ follow from (\ref{eq:c_pm_qp})
(with the subsequent normalization).
The operator of the electric external field
in the long-wave approximation reads
\begin{equation}
\hat{f}_{\lambda\mu} = e
\frac{1}{1+\delta_{\mu ,0}}
r^{\lambda} (Y_{\lambda \mu} +  Y^{\dag}_{\lambda \mu}) .
\end{equation}
Further, $e^{eff}_s$ is the effective charge (in the
dipole case $e^{eff}_p= N/A$ and $e^{eff}_n=-  Z/A$ and
$e$ is the proton charge);
 $\omega_{\nu}$ is the energy of the the RPA state $|\nu>$.

It is worth noting that, unlike the standard definition of the
strength function with using $\delta (\omega - \omega_{\nu})$, we
exploit here the Lorentz weight. It is convenient to simulate
smoothing effects.

For direct use of the expression (\ref{eq:strength_function}), we still have
to know all the RPA eigenvalues and eigenvectors. This needs an appreciable
computational effort. To avoid this, let's recast (\ref{eq:strength_function})
to the form which does not need the information on the particular RPA states
\cite{Malov}.
For this aim, we will use the Cauchy residue theorem. Namely, the
strength function will be recast as a sum of the residues for the poles
$z=\pm \omega_{\nu}$. Since the sum of all the residues (covering all
the poles) is zero, the residues with $z=\pm \omega_{\nu}$ (whose
calculation is time consuming) can be replaced by the sum of residues
with $z=\omega \pm i(\Delta /2)$ and $z=\pm \varepsilon_{ph}$ whose
calculation is much less expensive. Now let's consider this procedure step by step.

First, we  use (\ref{eq:c_pm_qp}) and rewrite the matrix element
(\ref{eq:tr_me}) to the form
\begin{eqnarray}
M_{E\lambda\mu \nu} &=&  - \frac{1}{\sqrt{2N_{\nu}}}\sum_s e^{eff}_s
\sum_{ij \epsilon {s}}
\frac{
<ij|f_{\lambda\mu}|0>}{\varepsilon_{ij}^2-\omega^2_{\nu}}
\\
&\cdot& \sum_{s'k'}
\{
\varepsilon_{ij} \bar{q}^{\nu}_{s'k'}
<ij|\hat{X}^s_{s'k'}|0>
+ i \omega_{\nu} \bar{p}^{\nu}_{s'k'}
<ij|\hat{Y}^s_{s'k'}|0>
\}
\nonumber\\
&=&- \frac{1}{\sqrt{2N_{\nu}}}
\sum_{s'k'}
\{
\bar{q}^{\nu}_{s'k'}A_{s'k'}^{(X)}(E\lambda\mu ) +
\bar{p}^{\nu}_{s'k'}A_{s'k'}^{(Y)}(E\lambda\mu )
\} \;
\end{eqnarray}
where
\begin{equation}\label{eq:A_X_1}
A_{s'k'}^{(X)}(E\lambda\mu ) =
\sum_s e^{eff}_s \sum_{ij \epsilon {s}}
\frac{\varepsilon_{ij}
      <ij|X^{s}_{s'k'}|0> <ij|f_{\lambda\mu}|0>}
      {\varepsilon_{ij}^2-\omega_{\nu}^2} \; ,
\end{equation}
\begin{equation}\label{eq:A_Y_1}
A_{s'k'}^{(Y)}(E\lambda\mu ) =
i \sum_s e^{eff}_s \sum_{ij \epsilon s}
\frac{\omega_{\nu}
       <ij|Y^{s}_{s'k'}|0><ij|f_{\lambda\mu}|0>}
      {\varepsilon_{ij}^2-\omega_{\nu}^2} \; .
\end{equation}
For the sake of brevity, we introduce the new index $ \beta = \{ skg \}$ where $g=1$ and 2
for time-even and time-odd quantities, respectively.
Then the
squared matrix element can be written in the compact form \cite{Malov,Kva_PRC}
\begin{equation} \label{Mel^2}
M_{E\lambda\mu }^2 = \frac{1}{2N}
\sum_{\beta \beta'} R_{\beta} R_{\beta'} A_{\beta} A_{\beta'}
=\frac{1}{2N} \sum_{\beta \beta'} 2N \frac{F_{\beta \beta'}}
  {\frac{\partial}{\partial \omega} \det D}
  A_{\beta} A_{\beta'}
= \sum_{\beta \beta'} \frac{F_{\beta \beta'}}
  {\frac{\partial}{\partial \omega} \det F} A_{\beta} A_{\beta'}
\end{equation}
where $F$ is the determinant of RPA matrix  (\ref{eq:X_c_qp4}),
$F_{\beta \beta'}$ is its algebraic supplement and
\begin{eqnarray}
R_{sk\; g=1} &=& \bar{q}^{\nu}_{sk}, \qquad\qquad R_{sk\; g=2}=\bar{p}^{\nu}_{sk},
\\
A_{sk\; g=1} &=& A_{sk}^{(X)}(E\lambda\mu ), \quad A_{sk\; g=2}= A_{sk}^{(Y)}(E\lambda\mu ).
\end{eqnarray}

Substituting (\ref{Mel^2}) to (\ref{eq:strength_function}), one gets
\begin{equation}\label{eq:str_func_2}
S_L(E\lambda\mu ; \omega)= \sum_{\nu}
\omega_{\nu}^{L}
\frac{\sum_{\beta \beta'} F^{\nu}_{\beta \beta'}A^{\nu}_{\beta} A^{\nu}_{\beta'} }
  {\frac{\partial}{\partial \omega_{\nu}} \det F^{\nu}} \zeta(\omega - \omega_{\nu}).
\end{equation}
Then, taking into account that the determinant $\det F^{\nu}$ has
first-order poles $\omega=\omega_{\nu}$, the strength function can be rewritten
through the residue of these poles on the complex plain, or equivalently,
through the corresponding contour integrals:
\begin{eqnarray}\label{eq:str_func_3}
S_L(E\lambda\mu ; \omega)&=& \sum_{\nu}
 Res \{ z^{L}
\frac{\sum_{\beta \beta'}F_{\beta \beta'}(z) A_{\beta}(z) A_{\beta'}(z) }
  {\det F(z)} \zeta(\omega - z) \}_{z=\omega_{\nu}}
 \nonumber  \\
 &=& \frac{1}{2\pi i}\sum_{\nu} \oint_{z=\omega_{\nu}} z^{L}
\frac{\sum_{\beta \beta'}F_{\beta \beta'}(z) A_{\beta}(z) A_{\beta'}(z) }
  {\det F(z)} \zeta(\omega - z) .
\end{eqnarray}
Unlike (\ref{eq:str_func_2}), the denominator in (\ref{eq:str_func_3}) includes
the RPA determinant instead of its derivative.

Following Cauchy theorem, sum of all the residues (covering all
possible poles of the strength function) is zero
and so one can express the residues with $z=\omega_{\nu}$ through
the rest of the others:
\begin{equation}
Res[S]_{z=\omega_{\nu}}= -(
Res[S]_{z=-\omega_{\nu}}+Res[S]_{z\to\infty}
+ Res[S]_{z=\omega \pm i(\Delta /2)}+Res[S]_{z=\pm\varepsilon_{ij}}
)
\end{equation}
where the poles $z=\omega \pm i(\Delta /2)$ and $z=\pm\varepsilon_{ij}$
originate  from  the Lorentz weight and denominator
$F_{\beta \beta'}(z) A_{\beta}(z) A_{\beta'}(z)$, respectively.
The RPA determinant $F(z)$ has zeros only at $z=\pm \omega_{\nu}$.

It's easy to prove that for $L =0,1,2$ we have
$\lim_{|z|\to \infty}S(\omega ,z)=0$. Also, $Res[S]_{z=-\omega_{\nu}}$ and
$Res[S]_{z=-\varepsilon_{ij}}$ can be neglected for large positive
$z$-values (high energies of giant resonances) and relevant values of
the averaging parameter $\Delta$. Remaining residues over the poles
$z=\omega \pm i(\Delta /2)$ and $z=\varepsilon_{ph}$
give the final outcome
\begin{equation}
Res[S]_{z=\omega_{\nu}} \simeq
- Res[S]_{z=\omega \pm i(\Delta /2)}- Res[S]_{z=\varepsilon_{ij}} .
\end{equation}

Finally, the strength function for $L=0,1,2$ reads
\begin{eqnarray}
\label{eq:sf_1}
S_L (E\lambda\mu ,\omega )
&=&\frac{1}{\pi} \Im
\left[
\frac{z^{L}\sum_{\beta \beta'}F_{\beta \beta'}(z) A_{\beta}(z) A_{\beta'}(z)}{F(z)}
\right]_{z=\omega +i(\Delta /2)}
\\
&+&
\sum_s (e^{eff}_s)^2 \sum_{ij \epsilon s}
\varepsilon_{ij}^{L}
<ij|f_{\lambda\mu}|0>^2 \zeta(\omega -\varepsilon_{ij}) .
\nonumber
\end{eqnarray}
The first term in (\ref{eq:sf_1})
is contributions of the residual interaction. It vanishes at $V_{\rm res}=0$.
The second term is the unperturbed (purely quasiparticle) strength function.

\subsection{General discussion}
\label{sec:general}

Equations (\ref{eq:X})-(\ref{eq:Y}),
(\ref{eq:kappa})-(\ref{eq:eta}),
(\ref{eq:c_pm_qp}),
 (\ref{eq:X_c_qp4}),
(\ref{eq:F_XX})-(\ref{eq:F_YY}), and
(\ref{eq:c_norm_F}) constitute the basic SRPA formalism.
Before proceeding the specification to Skyrme functional it is worth to
comment some essential points of the model.
\\
$\bullet$ One may show (e.g. by using a standard derivation of the
matrix RPA) that the separable Hamiltonian
(\ref{eq:H_sep}) with one-phonon states (\ref{eq:Q_op}) results in the
SRPA equations (\ref{eq:X_c_qp4})-(\ref{eq:F_YY}) if to express
unknowns $c^{\nu \pm}_{ph}$ through $\bar{q}_{\bar{k}}$ and
$\bar{p}_{\bar{k}}$.  Familiar RPA equations for unknowns $c^{\nu
  \pm}_{ph}$ require the RPA matrix of a high rank equal to size of
the $1ph$ basis. The separable approximation allows to reformulate the
RPA problem in terms of a few unknowns $\bar{q}_{\bar{k}}$ and
$\bar{p}_{\bar{k}}$ (see relation (\ref{eq:c_pm_qp})) and thus to
minimize the computational effort. As is seen from (\ref{eq:X_c_qp4}),
the rank of the SRPA matrix is equal to $4K$ (where $K$ is the number of the
separable operators) and hence is quite low.
\\
$\bullet$ The number of RPA states $|\nu>$ is equal to the number
of the relevant $1ph$ configurations used in the calculations. In
heavy nuclei, this number ranges the interval
$10^3$-$10^4$.  Every RPA state $|\nu>$ is characterized
by the particular set of the values $\bar{q}^{\nu}_{sk}$ and
$\bar{p}^{\nu}_{sk}$ which, following (\ref{eq:c_pm_qp}),
self-consistently regulate relative contributions of different time-even
and time-odd operators of the residual interaction to this state.
\\
$\bullet$ Eqs. (\ref{eq:X}), (\ref{eq:Y}), (\ref{eq:kappa}),
(\ref{eq:eta}) relate the basic SRPA values with the starting
functional and input operators $\hat{Q}_{sk}$ and $\hat{P}_{sk}$ by a
simple and physically transparent way.
After choosing  the initial operators $\hat{Q}_{sk}$,
all other SRPA values are straightforwardly
determined following the steps
\begin{equation}\label{eq:Q-initial}
\hat{Q}_{sk} \; \Rightarrow \;
\langle 0|[\hat{Q}_{sk},{\hat J}^{\alpha}_s]|0\rangle  \; \Rightarrow \;
\hat{Y}_{sk}, \; \eta_{sk,s'k'}^{-1} \; \Rightarrow \;
\hat{P}_{sk} \; \Rightarrow \;
\langle 0|[\hat{P}_{sk},{\hat J}^{\alpha}_s]|0\rangle  \; \Rightarrow \;
\hat{X}_{sk}, \; \kappa_{sk,s'k'}^{-1} .
\end{equation}
As is discussed in Sec. \ref{sec:Q_choice}, the proper choice of $\hat{Q}_{sk}$ is crucial to
achieve good convergence of the separable expansion (\ref{V_sep}) at
a minimal number of separable operators.
\\
$\bullet$
 SRPA restores the conservation laws (e.g. translational invariance)
violated by the static mean field. Indeed, let's assume a symmetry mode
with the generator $\hat{P}_{\rm sym}$. Then, to keep the conservation law
$[\hat{H},\hat{P}_{\rm sym}]=0$,  we simply have to include
$\hat{P}_{\rm sym}$ into the set of the input generators $\hat{P}_{sk}$
together with its complement $\hat{Q}_{\rm sym}=i[\hat{H},\hat{P}_{\rm sym}]$.
\\
$\bullet$ The basic SRPA operators can be expressed via
the separable residual interaction (\ref{V_sep}):
\begin{equation} \label{eq:XY_V_res}
\hat{X}_{sk} = -i [\hat{V}^{\rm sep}_{\rm res}, \hat{P}_{sk}]_{ph}, \qquad
\hat{Y}_{sk} = -i [\hat{V}^{\rm sep}_{\rm res}, \hat{Q}_{sk}]_{ph}
\end{equation}
where the index $ph$ means the $1ph$ part of the commutator. It is seen that
the time-odd operator $\hat{P}_{sk}$ retains the time-even part of
$V_{\rm res}^{\rm sep}$
to build $\hat{X}_{sk}$. Vice versa, the commutator with the time-even operator
$\hat{Q}_{sk}$ keeps the time-odd part of $V_{\rm res}^{\rm sep}$ to
build $\hat{Y}_{sk}$. Equations (\ref{eq:XY_V_res}) hints also the relation between
the SRPA operators and the true (not separable) residual two-body interaction.
\\
$\bullet$
Some of the SRPA values read as averaged commutators between
time-odd and time-even operators. This allows to establish useful relations
with other models. For example, (\ref{eq:kappa}), (\ref{eq:eta}) and
(\ref{eq:XY_V_res}) give
\begin{eqnarray}
\label{eq:kappa_V}
\kappa_{s'k',sk}^{-1 } = - i \langle 0|[\hat{P}_{s'k'},{\hat X}_{sk}^{s'}]|0\rangle  =
-  \langle 0|[\hat{P}_{s'k'},[\hat{V}_{\rm res}^{\rm sep},{\hat P}_{sk}]]|0\rangle  ,
\\
\label{eq:eta_V}
\eta_{s'k',sk}^{-1 } = -i \langle 0|[\hat{Q}_{s'k'},{\hat Y}_{sk}^{s'}]|0\rangle  =
- \langle 0|[\hat{Q}_{s'k'},[\hat{V}_{\rm res}^{\rm sep},{\hat Q}_{sk}]]|0\rangle  .
\end{eqnarray}
Similar double commutators (but with the full Hamiltonian instead
of $\hat{V}_{\rm res}^{\rm sep}$) correspond to $m_3$ and $m_1$ sum rules,
respectively, and so represent the spring and inertia parameters
\cite{Re_PRA_90} in the basis of collective generators $\hat{Q}_{sk}$
and $\hat{P}_{sk}$. This demonstrates the connection of the SRPA with
the sum rule approach \cite{BLM_79,LS_89} and local RPA \cite{Re_PRA_90}.
\\
$\bullet$
The commutator form of the  SRPA values allows to represent them
through the matrix elements from the operators entering the commutators.
Namely, the strength constants and responses gain the form
(see Appendix \ref{sec:rep_me})
\begin{eqnarray}\label{eq:com_Q}
 i<0|[\hat{Q}_{sk},\hat{B}_s]|0>
&=&  -4  \sum_{ij}^{K_i,K_j>0}
 <ij|\hat{Q}_{sk}|0> \; \Im\{<ij|\hat{B}_s|0>\} \; ,
\\
\label{eq:com_P}
 i<0|[\hat{P}_{sk},\hat{B}_s]|0>
&=& -4i  \sum_{ij}^{K_i,K_j>0}
 <ij|\hat{P}_{sk}|0> \; \Re\{<ij|\hat{B}_s|0>\} \;
\end{eqnarray}
where $\hat B$ equals to  $\hat X$
for the strength constants and to $\hat J$ (the density operator) for the responses.
Since the involved matrix elements are already in our disposal,
these forms considerably simplify the calculations. Besides, these forms
are convenient for analysis. For example, they allow to determine
the conditions of vanishing some response components,
see Appendix \ref{sec:rep_me} for more details.
\\
$\bullet$ In fact, SRPA is the first TDHF iteration with the initial wave
function (\ref{eq:scaling}).
A single iteration is generally not enough to get the
complete convergence of TDHF results. However, SRPA calculations demonstrate
that high accuracy can be achieved even in this case if to ensure the optimal
choice of the input operators  $\hat{Q}_{sk}$ and $\hat{P}_{sk}$ and keep
sufficient amount of the separable terms (see discussion in Sec.
\ref{sec:Q_choice}). In this case, the first iteration already gives quite
accurate results.
\\
$\bullet$ SRPA equations are very general and, after simple modifications,
can be applied to diverse systems (atomic nuclei, atomic clusters, etc.)
described by density and current-dependent functionals, see for the case
of  atomic clusters Refs.
\cite{Ne_PRA_98,Ne_EPJD_98,Ne_PRL_sciss,Ne_AP_02,Ne_EPJD_02,Ne_PRA_04}.
Even Bose systems can be covered if one redefines
the many-body wave function (\ref{Psi_Thouless}) exhibiting the perturbation
through the elementary excitations. In this case, the Slater determinant for
1ph excitations should be replaced by a perturbed many-body function in terms
of elementary bosonic excitations.

\section{Specification for Skyrme functional}
\label{sec:skyrme}
\subsection{Skyrme functional and its mean field Hamiltonian}

We use the Skyrme  functional \cite{Skyrme} in the particular form
\cite{Engel_75,Re_AP_92,Dob}
\begin{equation}
   {E}  =  \int d{\vec r}\left({\cal H}_{\rm kin}
 +{\cal H}_{\rm Sk}(\rho_s,\tau_s,
   \vec{\sigma}_s,\vec{j}_s,\vec{J}_s)
                           + {\cal H}_{\rm C}(\rho_p) \right) - E_{cm},
\end{equation}
where
\begin{eqnarray}
   {\cal H}_{\rm kin} &= &  \frac{\hbar^2}{2m} \tau ,
 \label{Ekin}
 \\
 {\cal H}_{\rm C}
 & = & \frac{e^2}{2}
 \int  d{\vec r}' \rho_p(\vec{r})
              \frac{1}{|\vec{r}-\vec{r}'|} \rho_p(\vec{r}')
         -\frac{3}{4} e^2(\frac{3}{\pi})^\frac{1}{3}
                          [ \rho_p(\vec{r})]^\frac{4}{3} ,
\label{Ecoul}
\\
   {\cal H}_{\rm Sk} &= &
                 \frac{b_0}{2}  \rho^2
                  -\frac{b'_0}{2} \sum_s \rho_s^2
    -\frac{b_2}{2} \rho (\Delta \rho)
     +\frac{b'_2}{2} \sum_s \rho_s (\Delta \rho_s)
    \\
        & &
          + \frac{b_3}{3}  \rho^{\alpha +2}
          - \frac{b'_3}{3} \rho^\alpha   \sum_s\rho_s^2
            \nonumber \\
     & &
     +  b_1 (\rho \tau - \vec{j}^2)
     - b'_1 \sum_s (\rho_s \tau_s - \vec{j}_s^2)
\nonumber\\
     & &
     - b_4 \left( \rho (\vec{\nabla}\vec{{\Im}})
      + \vec{\sigma} \cdot (\vec{\nabla} \times \vec{j})\right)
     -b'_4 \sum_s \left( \rho_s(\vec{\nabla} \vec{\Im}_s)
              + \vec{\sigma}_s \cdot (\vec{\nabla} \times \vec{j}_s) \right)
\nonumber
 \label{Esky}
\end{eqnarray}
are kinetic, Coulomb and Skyrme terms respectively. The densities and currents
used in this functional are defined in the Appendix {\ref{sec:dens_curr}.
Densities without the index $s$ involve both neutrons and protons, e.g.
$\rho=\rho_p+\rho_n$. Parameters $b$ and $\alpha$ are fitted to describe ground
state properties of atomic nuclei. The value $E_{cm}$ is the center mass correction.

First functional derivatives read \cite{Re_AP_92}
\begin{eqnarray}\label{U_s}
U_s ({\vec r}) = \frac{\delta E}{\delta \rho_s ({\vec r})}
&=& b_0\rho ({\vec r}) - b_0'\rho_s ({\vec r})
\\
&+& b_3 \frac{\alpha +2}{3} \rho^{\alpha +1} ({\vec r})
  - \frac{b_3'}{3} \{ \alpha\rho^{\alpha -1} ({\vec r})
     \sum_{s'}\rho^2_{s'}({\vec r}) + 2\rho^{\alpha} ({\vec r}) \rho_s ({\vec r}) \}
\nonumber \\
&+& b_1\tau ({\vec r}) - b_1'\tau_s ({\vec r})
 - b_2 \Delta_{\vec r} \; \rho ({\vec r}) + b_2' \Delta_{\vec r} \; \rho_s ({\vec r})
\nonumber \\
  &-& b_4 {\vec \nabla}_{\vec r} \; {\vec \Im} ({\vec r})
  - b_4'{\vec \nabla}_{\vec r} \; {\vec \Im}_s ({\vec r})
\nonumber \\
  &+& \delta_{s,p} \; e^2 \{
    \int d{\vec r_1} \frac{\rho_p({\vec r_1})}{|{\vec r}-{\vec r_1}|}
  - (\frac{3}{\pi})^{1/3}[\rho_p({\vec r})]^{1/3}
    \} ,
\nonumber\\ \label{B_s}
B_s ({\vec r}) = \frac{\delta E}{\delta \tau_s({\vec r})}
&=& \frac{\hbar^2}{2m} + b_1 \rho ({\vec r}) -b_1' \rho_s ({\vec r}),
\\ \label{W_s}
{\vec W}_s ({\vec r}) = \frac{\delta E}{\delta {\vec \Im}_s({\vec r})}
&=& b_4 {\vec\nabla}_{\vec r} \; \rho ({\vec r}) + b_4' {\vec\nabla}_{\vec r} \; \rho_s ({\vec r}).
\end{eqnarray}
for time-even densities and currents and
\begin{eqnarray}\label{A_s}
{\vec A}_s ({\vec r}) &=& \frac{\delta E}{\delta {\vec j}_s({\vec r})}
           = - 2b_1 {\vec j} ({\vec r}) + 2b_1' {\vec j}_s ({\vec r})
             - b_4 (\vec{\nabla}_{\vec r} \times \vec{\sigma} ({\vec r}))
             - b_4' (\vec{\nabla}_{\vec r} \times \vec{\sigma}_s ({\vec r}))\;,
\\\label{S_s}
{\vec S}_s ({\vec r}) &=& \frac{\delta E}{\delta {\vec \sigma}_s({\vec r})}
           = - b_4 (\vec{\nabla}_{\vec r} \times \vec{j} ({\vec r}))
           - b_4'(\vec{\nabla}_{\vec r} \times  \vec{j}_s ({\vec r}))\;.
\end{eqnarray}
for time-odd ones. The last line in (\ref{U_s}) is the Coulomb contribution. It includes only
proton density ($s=p$). Since static ground-state time-odd densities are zero, the values
${\vec \sigma}({\vec r})$ and ${\vec j}({\vec r})$ in (\ref{A_s}) and (\ref{S_s})
are reduced to the time-dependent density variations (\ref{eq:dens_J(t)}).
We assume that the functional derivatives  (\ref{U_s})-(\ref{S_s}) and the
involved densities are have time dependent
but, for the sake of simplicity, do not depict this.

Using (\ref{U_s})-(\ref{W_s}), we get
the time-even part of the mean field Hamiltonian (\ref{eq:h_0})
\begin{eqnarray}\label{h_mean_field_Rein}
\hat{h}^s_{+}(\vec r) &=&
U_s (\rho,z)-\vec{\nabla} B_s(\rho,z) {\vec \nabla}
-i\vec{W}_s(\rho,z) \cdot {\vec \nabla} \times \hat{\vec \sigma}
\\
&=&
U_s (\rho,z)
+\overleftarrow{\nabla} B_s(\rho,z)  {\vec \nabla}
-i \vec{W}_s(\rho,z) \cdot {\vec \nabla} \times \hat{\vec \sigma} .
\label{h_mean_field_wibp}
\nonumber
\end{eqnarray}
If to implement the static densities and currents, one gets the
ground-state Hamiltonian, i.e.  $\hat{h}^s_{+} \to \hat{h}^s_{0}$.
The ground-state mean filed of even-even axial nuclei has no
any contributions from the time-odd variables.

The part of the mean field Hamiltonian (\ref{eq:h_0}) from the time-odd densities
reads
\begin{eqnarray}\label{h_response_Rein}
\hat{h}^s_{-}(\vec r) &=& -\frac{i}{2} \{\vec {A}_s(\rho,z),\vec{\nabla}\}
                        +\frac{1}{2}\{\vec {S}_s(\rho,z),{\hat {\vec \sigma}}\}
\\
\label{h_response_wibp}
&=& \frac{i}{2} (\overleftarrow{\nabla} \cdot \vec{A}_s(\rho,z)-\vec {A}_s(\rho,z)\cdot \vec{\nabla})
                        +\vec {S}_s(\rho,z)\cdot {\hat {\vec \sigma}} .
\nonumber
\end{eqnarray}
It is used only for derivation of the response Hamiltonian.
Hence the involved current and spin densities
are represented merely by their time-dependent variations.



\subsection{Time-even response}

For time-even densities, the second functional derivatives read
\begin{eqnarray}\label{dE_drho_drho}
\frac{\delta^2 E}{\delta \rho_{s_1} ({\vec r}_1)\delta \rho_s ({\vec r})}
&=& \frac{\delta U_s ({\vec r})}{\delta \rho_{s_1} ({\vec r}_1)}
\\
&=& \{
    b_0 - b_0'\delta_{ss'} +(- b_2 + b_2'\delta_{ss_1})\Delta_{{\vec r}_1}
   + b_3 (\alpha +2)(\alpha +1) \rho^{\alpha} ({\vec r})
\nonumber\\
&-&  b_3' [ \frac{(\alpha (\alpha -1)}{3} \rho^{\alpha -2}({\vec r}) \sum_{s_2}\rho^2_{s_2}({\vec r})
          + \frac{2\alpha}{3} \rho^{\alpha -1}({\vec r}) (\rho_s ({\vec r}) + \rho_{s_1}({\vec r}))
          + \delta_{ss_1}\frac{2}{3}\rho^{\alpha} ({\vec r}) ]
\nonumber \\
&-& \delta_{s,s_1} \delta_{s,p}\frac{1}{3}(\frac{2}{\pi})^{1/3}[\rho_p({\vec r})]^{-2/3}
     \} \;
    \delta ({\vec r}-{\vec r}_1)
\nonumber \\
&+& \delta_{s,s_1} \delta_{s,p} \;
\frac{e^2}{|{\vec r}-{\vec r_1}|} \; ,
\nonumber\\ \label{dE_drho_dtau}
\frac{\delta^2 E}{\delta \tau_{s_1} ({\vec r}_1)\delta \rho_s ({\vec r})}
&=& \frac{\delta U_s ({\vec r})}{\delta \tau_{s_1} ({\vec r}_1)}
 = [ b_1 - b_1'\delta_{ss'} ]\delta ({\vec r}-{\vec r}_1) \; ,
\\ \label{dE_drho_dJ}
\frac{\delta^2 E}{\delta {\vec \Im}_{s_1} ({\vec r}_1)\delta \rho_s ({\vec r})}
&=& \frac{\delta U_s ({\vec r})}{\delta {\vec \Im}_{s_1} ({\vec r}_1)}
 = [ b_4 + b_4'\delta_{ss_1} ]{\vec \nabla}_{{\vec r}_1} \delta ({\vec r}-{\vec r}_1)
\; ,
\\
 \label{dE_dJ_drho}
\frac{\delta^2 E}{\delta \rho_{s_1} ({\vec r}_1)\delta {\vec \Im}_{s} ({\vec r})}
&=&
\frac{\delta \vec{W}_{s} ({\vec r})}{\delta \rho_{s_1} ({\vec r}_1)}
 = -[ b_4 + b_4'\delta_{ss_1} ]{\vec \nabla}_{{\vec r}_1} \delta ({\vec r}-{\vec r}_1) \; .
\end{eqnarray}
The last two terms in (\ref{dE_drho_drho}) represent the exchange and direct Coulomb
contributions. The pairing second functional derivative is not presented here.
The pairing contribution to the response is considered in Sec. \ref{sec:pairing}.

Following (\ref{eq:X}), the operator $\hat{X}_{s k}^{s'}({\vec r})$ reads
\begin{eqnarray}\label{eq:X_expl_int}
\hat{X}_{s_1 k_1}^{s}({\vec r}) &=&
i\sum_{\alpha \alpha_1} \int d{\vec r}_1
[\frac{\delta^2 {\cal E}}{\delta J_{s_1}^{\alpha_1}({\vec r_1})
\delta J_{s}^{\alpha}({\vec r})}]_{J=\bar{J}}
<0|[\hat{P}_{s_1k_1} ,{\hat J}_{s_1}^{\alpha_1}]|0> ({\vec r}_1)
{\hat J}_{s}^{\alpha}({\vec r})]
\nonumber\\
&=&
U_{s_1 k_1}^{s}(\vec r) \hat{\rho}_{s}(\vec r)+
B_{s_1 k_1}^{s}(\vec r) \hat{\tau}_{s}(\vec r)  +
\vec{W}_{s_1 k_1}^{s}(\vec r)\hat{\vec{\Im}}_{s}(\vec r)
+ \delta_{s,s_1}\hat{X}_{s k_1}^{s(\rm pair)}
\label{eq:X_expl_int_2}
\end{eqnarray}
and its matrix element is
\begin{eqnarray}\label{eq:<iX|j>}
<ij|\hat{X}_{s_1 k_1}^{s}|\tilde 0> &=&  \int d\vec{r}
\{
U_{s_1 k_1}^{s}(\vec r) \rho^{s}_{ij}(\vec r) +
B_{s_1 k_1}^{s}(\vec r) \tau^{s}_{ij}(\vec r) +
\vec{W}_{s_1 k_1}^{s}(\vec r) \vec{\Im}^{s}_{ij}(\vec r)\}
\\
&+& \delta_{s,s_1}<ij|\hat{X}_{s k_1}^{s(pair)}|\tilde 0>
\nonumber
\end{eqnarray}
where $\rho^{s}_{ij}(\vec r)$, $\tau^{s}_{ij}(\vec r)$, and
$\vec{\Im}^{s}_{ij}(\vec r)$ are transition densities
(\ref{trans_dens_rho})-(\ref{trans_dens_Im}). Expressions for operator
$\hat{X}_{s k_1}^{s(\rm pair)}$ and matrix element
$<ij|\hat{X}_{s k_1}^{s(pair)}|\tilde 0>$ are done in (\ref{X_pair})
and the Appendix \ref{sec:matr_el_X}, respectively.
The functions $U_{s_1 k_1}^{s}(\vec r)$,
$B_{s_1 k_1}^{s}(\vec r)$, and $\vec W_{s_1 k_1}^{s}(\vec r)$ read
\begin{eqnarray}\label{U_X_k}
U_{s_1 k_1}^{s}({\vec r})
&=& \int d {\vec r}_1 \;
 [ \frac{\delta^2 E}{\delta \rho_{s_1} ({\vec r}_1)\delta \rho_{s} ({\vec r})}
\rho_{X, s_1  k_1}({\vec r}_1)
+
\frac{\delta^2 E}{\delta \tau_{s_1} ({\vec r}_1)\delta \rho_{s} ({\vec r})}
\tau_{X, s_1 k_1}({\vec r}_1)
\\
&& \qquad \; +
\frac{\delta^2 E}{\delta \vec{\Im}_{s_1} ({\vec r}_1)\delta \rho_{s} ({\vec r})}
\vec{\Im}_{X, s_1 k_1}({\vec r}_1)]
\nonumber\\
&=& {\bar U}_{s_1 k_1}^{s}(\rho,z) \cos \mu\theta ,
\nonumber\\
\label{B_X_k}
 B_{s_1 k_1}^{s}({\vec r})&=&
 \int d {\vec r}_1 \;
 \frac{\delta^2 E}{\delta \rho_{s_1} ({\vec r}_1)\delta \tau_{s} ({\vec r})}
\rho_{X, s_1 k_1}({\vec r}_1) =
{\bar B}_{s_1 k_1}^{s}(\rho,z) \cos \mu\theta ,
\\
\label{W_X_k}
 \vec{W}_{s_1 k_1}^{s}({\vec r})
&=& \int d {\vec r}_1 \;
[ \frac{\delta^2 E}{\delta \rho_{s_1} ({\vec r}_1)\delta \vec{\Im}_{s} ({\vec r})}
\rho_{X, s_1 k_1}({\vec r}_1)
+
\frac{\delta^2 E}{\delta \tau_{s_1} ({\vec r}_1)\delta \vec{\Im}_{s} ({\vec r})}
\tau_{X, s_1 k_1}({\vec r}_1)
\\
&& \qquad \; +
\frac{\delta^2 E}{\delta \vec{\Im}_{s_1} ({\vec r}_1)\delta \vec{\Im}_{s} ({\vec r})}
\vec{\Im}_{X, s_1 k_1}({\vec r}_1)]
\nonumber\\&=&
{\vec e}_{\rho}{\bar W}_{s_1 k_1; \rho}^{s}(\rho,z)\cos\mu\theta
+{\vec e}_z{\bar W}_{s_1 k_1; z}^{s}(\rho,z)\cos\mu\theta
+{\vec e}_{\theta}{\bar W}_{s_1 k_1; \theta}^{s}(\rho,z) \sin\mu\theta
\nonumber
\end{eqnarray}
where the values ${\bar U}_{s_1 k_1}^{s}(\rho,z)$, ${\bar B}_{s_1
k_1}^{s}(\rho,z)$, ${\bar W}_{s_1 k_1;\;\rho}^{s}(\rho,z)$,
${\bar W}_{s_1 k_1;\;z}^{s}(\rho,z)$, and
${\bar W}_{s_1 k_1;\;\theta}^{s}(\rho,z)$
are given in the Appendix \ref{sec:matr_el_X}.

Contribution of the Coulomb integral to the matrix element
(\ref{eq:<iX|j>}) reads
\begin{eqnarray}\label{eq:<iX|j>_coul}
<i|\hat{X}_{s_1 k_1}^{s}|j>_{Coul} &=&
\delta_{s_1,p}\delta_{s,p} e^2 \int d{\vec r}\int d{\vec r}_1
\frac{\rho_{X_{k_1},p}(\rho_1,z_1)\cos\mu\theta_1}
     {|\vec{r}-\vec{r}_1|}\rho_{ij}^{s}({\vec r})
\\
&=&
\delta_{s_1,p}\delta_{s,p} e^2 \int d{\vec r} \;
U^{\rm Coul}_{k_1} (\rho,z) \cos \mu\theta \rho_{ij}^{s}({\vec r})
\;
\nonumber
\end{eqnarray}
where $U^{\rm Coul}_{k_1} (\rho,z)$ is determined in (\ref{coul_rho_z}).
It should be included to $\bar U_{s_1 k_1}^{s}(\rho, z)$, see (\ref{bar_U_k}).

In our case, the singularity problem for the Coulomb interaction
cannot be treated by the familiar methods.  These methods mainly deal
with spherical systems \cite{FFT} or axial systems in the static ground state
\cite{Stoitsov_Dob} and, in any case, do not master the specific structure of
the dynamical response in axial systems, given in (\ref{eq:<iX|j>_coul}).
Hence we propose in Appendix
\ref{sec:coul_inter} the new prescription.

\subsection{Time-odd response}

For time-odd densities, the second functional derivatives read
\begin{eqnarray}\label{dE_dj_dj}
\frac{\delta^2 E}{\delta \vec{j}_{s_1} ({\vec r}_1)\delta \vec{j}_{s} ({\vec r})}
&=& \frac{\delta \vec{A}_{s} ({\vec r})}{\delta \vec{j}_{ s_1} ({\vec r}_1)}
 = 2 [- b_1 + b_1' \delta_{s,s_1}] \delta ({\vec r}_1-{\vec r})\; ,
\\ \label{dE_dj_dsigma}
\frac{\delta^2 E}{\delta \vec{\sigma}_{s_1} ({\vec r}_1)\delta \vec{j}_{s} ({\vec r})}
&=& \frac{\delta \vec{A}_{s} ({\vec r})}{\delta \vec{\sigma}_{s_1} ({\vec r}_1)}
 = -[ b_4 + b_4' \delta_{s,s_1}] \vec{\nabla}_{{\vec r}_1} \times \delta ({\vec r}_1-{\vec r}) \; ,
\\ \label{dE_dsigma_dj}
\frac{\delta^2 E}{\delta \vec{j}_{s_1} ({\vec r}_1)\delta \vec{\sigma}_{s} ({\vec r})}
&=& \frac{\delta \vec{S}_{s} ({\vec r})}{\delta \vec{j}_{s_1} ({\vec r}_1)}
 = - [ b_4 + b_4' \delta_{s,s_1}] \vec{\nabla}_{{\vec r}_1} \times \delta ({\vec r}_1-{\vec r}) \; .
\end{eqnarray}
 Following (\ref{eq:Y}), the operator $\hat{Y}_{s_1 k_1}^{s'}({\vec r})$ has the form
\begin{eqnarray}\label{eq:X_expl_int}
\hat{Y}_{s_1 k_1}^{s}({\vec r}) &=&
i\sum_{\alpha \alpha_1} \int d{\vec r}_1
[\frac{\delta^2 {\cal E}}{\delta J_{s_1}^{\alpha_1}({\vec r_1})
\delta J_{s}^{\alpha}({\vec r})}]_{J=\bar{J}}
<0|[\hat{Q}_{s_1k_1} ,{\hat J}_{s_1}^{\alpha_1}]|0> ({\vec r}_1)
{\hat J}_{s}^{\alpha}({\vec r})]
\nonumber\\
&=&
\vec{A}_{s_1 k_1}^{s}(\vec r) \hat{\vec{j}}_{s}(\vec r)+
\vec{S}_{s_1 k_1}^{s}(\vec r) \hat{\vec{\sigma}}_{s}(\vec r) \; .
\end{eqnarray}
and its matrix element is
\begin{equation}\label{Y_matr_el_gen}
<ij|\hat{Y}_{s_1 k_1}^{s}|\tilde 0> = \int d\vec r \;
[{\bar A}_{s_1 k_1 ;\rho}^{s}(\vec r) j^{s}_{ij; \rho}(\vec r)
+{\bar A}_{s_1 k_1 ;z}^{s}(\vec r) j^{s}_{ij; z}(\vec r)
+{\bar S}_{s_1 k_1 ;\theta}^{s}(\vec r) s^{s}_{ij; \theta}(\vec r)]
\end{equation}
where
\begin{eqnarray}
\vec{A}_{s_1k_1}^{s}(\vec r)&=& \int d{\vec r}_1 \;
[ \frac{\delta^2 E}{\delta \vec{j}_{s_1} ({\vec r}_1)\delta \vec{j}_{s} ({\vec r})}
\vec{j}_{Y, s_1k_1}
+
\frac{\delta^2 E}{\delta \vec{\sigma}_{s_1} ({\vec r}_1)\delta \vec{j}_{s} ({\vec r})}
\vec{s}_{Y, s_1k_1} ]
\nonumber\\
&=& 2 [- b_1 + b_1' \delta_{s,s_1}] \vec{j}_{Y, s_1k_1}
- [b_4 + b_4' \delta_{s,s_1}] (\vec{\nabla} \times \vec{s}_{Y, s_1k_1})
\nonumber\\
&=& {\vec e}_{\rho}{\bar A}_{s_1k_1; \rho}^{s}(\rho,z)\cos\mu\theta
+{\vec e}_z{\bar A}_{s_1k_1 ;z}^{s}(\rho,z)\cos\mu\theta
+{\vec e}_{\theta}{\bar A}_{s_1k_1 ;\theta}^{s}(\rho,z)\sin\mu\theta \:,
\label{A_Y_k}
\end{eqnarray}
\begin{eqnarray}
\vec{S}_{s_1k_1}^{s}(\vec r)&=& \int d{\vec r}_1 \;
\frac{\delta^2 E}{\delta \vec{j}_{s_1} ({\vec r}_1)\delta \vec{\sigma}_{s} ({\vec r})}
\vec{j}_{Y, s_1k_1}
\nonumber\\
&=& -[ b_4 + b_4' \delta_{s,s_1}](\vec{\nabla} \times \vec{j}_{Y, s_1k_1})
\nonumber\\
&=& {\vec e}_{\rho}{\bar S}_{s_1k_1; \rho}^{s}(\rho,z)\sin\mu\theta
+{\vec e}_z{\bar S}_{s_1k_1; z}^{s}(\rho,z)\sin\mu\theta
+{\vec e}_{\theta}{\bar S}_{s_1k_1; \theta}^{s}(\rho,z)\cos\mu\theta
\label{S_Y_k} \; ,
\end{eqnarray}
and
\begin{eqnarray}
(\vec{\nabla} \times \vec{s}_{Y, sk}) &=&
 {\vec e}_{\rho}
   [\frac{\mu}{\rho}s_{Y, sk ;z}(\rho,z)-\partial_z s_{Y, sk; \theta}(\rho,z)]\cos\mu\theta
\\
&+& {\vec e}_{z}
   [\frac{1}{\rho}\partial_{\rho}(\rho s_{Y, sk; \theta})
   -\frac{\mu}{\rho}s_{Y, sk; \rho}]\cos\mu\theta
+ {\vec e}_{\theta}
   [\partial_z s_{Y, sk; \rho}-\partial_{\rho} s_{Y, sk; z}]\sin\mu\theta \:,
\nonumber \\
(\vec{\nabla} \times \vec{j}_{Y, sk}) &=&
 {\vec e}_{\rho}
 [-\frac{\mu}{\rho}j_{Y, sk; z}(\rho,z)-\partial_z j_{Y, sk; \theta}(\rho,z)]\sin\mu\theta
\\
&+& {\vec e}_{z}
 [\frac{1}{\rho}\partial_{\rho}(\rho j_{Y, sk; \theta})
 +\frac{\mu}{\rho}j_{Y, sk; \rho}]\sin\mu\theta
+ {\vec e}_{\theta}[\partial_z j_{Y, sk; \rho}
-\partial_{\rho} j_{Y, sk; z}]\cos\mu\theta \; .
\nonumber
\end{eqnarray}
Expressions for
${\bar A}_{Y, sk; \rho}^{s'}(\rho,z)$,
${\bar A}_{Y, sk; z}^{s'}(\rho,z)$,
${\bar A}_{Y, sk; \theta}^{s'}(\rho,z)$,
${\bar S}_{Y, sk; \rho}^{s'}(\rho,z)$,
${\bar S}_{Y, sk; z}^{s'}(\rho,z)$, and
${\bar S}_{Y, sk; \theta}^{s'}(\rho,z)$ are given in the Appendix
\ref{sec:matr_el_Y}.

\section{Choice of initial operators}
\label{sec:Q_choice}

As was discussed in Sec. \ref{sec:general}, all SRPA operations
start from initial (generating) operators $\hat{Q}_{sk}$,
see the sequence of the model steps in  (\ref{eq:Q-initial}).
The SRPA formalism itself does not provide
these operators. At the same time, their proper choice
is crucial to get good convergence of the separable expansion
(\ref{V_sep}) with a minimal number of separable operators.
The choice should be simple and universal in the sense that it can
be applied equally well to all modes and excitation channels.

We propose the choice inspired by physical arguments.
The main idea is that the generating operators should explore
different spatial regions of the nucleus, the surface as well
as the interior.
The leading scaling generator should have the form of
the applied external field in the long-wave approximation,
which is most sensitive to the surface of the system. Since
nuclear collective motion dominates in the surface region,
already this generator should provide a good description.
Next generators should be localized more in the interior
to describe an interplay of surface and volume vibrations.
For $E\lambda$-excitations in spherical nuclei, the best set of the generators
was found to be \cite{srpa_PRC_02}
\begin{equation}
  \hat{Q}_{k}({\vec r})
  =
  R_{k}(r)(Y_{\lambda\mu} (\Omega )+Y^*_{\lambda\mu} (\Omega ))
  \quad,\quad
  \hat{P}_{k}
  =
  i[\hat{H},\hat{Q}_{k}]
\label{eq:scale_op}
\end{equation}
with
\begin{equation}
  R_{k}(r)
  =
  \left\{
  \begin{array}{ll}
   r^{\lambda }, & k\!=\!1
  \\
  j_{\lambda}(q^k_{\lambda}r), & k\!=\!2,3,4
\\
\end{array}
\right.
\label{eq:actualset}
\end{equation}
$$
  q^{k}_{\lambda} = a_k\frac{z_{\lambda}}{R_{\rm diff}} ,\quad
  a_2\!=\!0.6\;,
  a_3\!=\!0.9\;,
  a_4\!=\!1.2
$$
where $R_{\rm diff}$ is the diffraction radius of the actual nucleus
and $z_{\lambda}$ is the first root in $j_{\lambda}(z_{\lambda})=0$.
The separable operators $\hat X_k$ and  $\hat X_k$ with $k=1$ are
mainly localized at the nuclear surface while the operators with $k>1$  are
localized more and more in the interior.  This simple set seems to be
a best compromise for the description of nuclear giant resonances in
light and heavy nuclei.

In deformed nuclei, we exploit even simpler set with
\begin{equation}\label{eq:radial}
  R_{k}(r) = r^{\lambda + 2(k-1)} .
\end{equation}
Similar to the previous set, the separable operators $\hat X_k$ and  $\hat X_k$
with $k=1$ are
mainly localized at the nuclear surface while the next ones are
localized more and more in the interior.
We expect that already the first generators with $k=1$ and 2 are quite enough
for description of giant resonances.

In the case of magnetic modes, the initial generators
should be the time-odd.  Following the same logic, the leading (k=1) operator
should coincide with the transition operator for
the external magnetic field in the long-wave approximation. The next generators
can be produced following the prescription (\ref{eq:radial}) for the radial
parts. The time-even conjugates of the time-odd generators can be obtained
as $\hat{Q}_{k} = i[\hat{H},\hat{P}_{k}]_{ph}$.

\section{Conclusions}
\label{sec:summary}

A general procedure for self-consistent factorization of the residual nuclear
interaction is proposed for arbitrary  density- and current-dependent
functionals.  The separable RPA (SRPA) constructed in the framework of this
approach can dramatically simplify the calculations while keeping high accuracy
of numerical results. The economical effect of SRPA is especially actual
 for deformed nuclei.
In the present contribution, SRPA is specified
for description of axial nuclei with Skyrme forces.

SRPA  can be used for description of $E\lambda$ (and $M\lambda$)
response in both spherical and deformed nuclei. The approach can
also serve for getting the basis of one-phonon RPA states for
further description of anharmonic corrections, vibrational admixtures
in the low-energy states in odd and odd-odd nuclei, etc.
One of the most promising lines of future studies is
dynamics of exotic nuclei obtained in radioactive beams.

\vspace{0.2cm} {\bf Acknowledgments.} The work was partly supported  by DFG
(project GZ:436 RUS 17/104/05), Bundesministerium f$\ddot u$r Bildung und
Forschung (project 06 DD 119), and  Heisenberg-Landau grant
(Germany-BLTP JINR, 2005).

\appendix 
\section{Cylindrical coordinates}
\label{sec:cylindr}
\subsection{Expressions for cylindrical coordinates}

Cylindrical coordinates are defined as
$$
x=\rho cos\vartheta \qquad
\vec e_x= \vec e_{\rho} cos\vartheta - \vec e_{\vartheta} sin\vartheta
\qquad \vec e_{\rho}  = \vec e_x cos\vartheta + \vec e_y sin\vartheta
$$
$$
y=\rho sin\vartheta \qquad
\vec e_y= \vec e_{\rho} sin\vartheta + \vec e_{\vartheta} cos\vartheta
\qquad \vec e_{\vartheta} = -\vec e_x sin\vartheta + \vec e_y cos\vartheta
$$
$$
z=z \hspace{4cm} \vec e_z=\vec e_z \hspace{4cm} \vec e_z=\vec e_z
$$
The gradient operator reads
$$
\vec{\nabla}
= \vec{e}_{\rho} \, \nabla_{\rho} + \vec{e}_{\vartheta} \, \nabla_{\vartheta}
+ \vec{e}_z \nabla_z
$$
where
$$
\nabla_{\rho} = \partial_{\rho}, \quad
\nabla_{\vartheta} =
\frac{1}{\rho} \partial_{\vartheta} = \frac{i \hat{L}_z}{\rho}, \quad
\nabla_z = \partial_z
$$
and $\hat L_z = -i \partial_{\vartheta}$ is the third component of the
orbital momentum operator.

The Laplacian, divergence, and curl read \cite{Korn}
$$
\Delta = \vec{\nabla} \cdot \vec{\nabla} =
\partial^2_{\rho} + \frac{1}{\rho} \partial_{\rho}
+ \frac{1}{\rho^2} \partial^2_{\vartheta} + \partial^2_z ,
$$
$$
div\vec{A} = \vec{\nabla} \cdot \vec{A} =
\frac{1}{\rho} \partial_{\rho}(\rho A_{\rho}) +
\frac{1}{\rho} \partial_{\vartheta} A_{\vartheta} +
\partial_z A_z ,
$$
$$
rot\vec{A} = [ \frac{1}{\rho} \partial_{\vartheta} A_z
   - \partial_z A_{\vartheta}] \vec{e}_{\rho} +
[ \partial_z A_{\rho} - \partial_{\rho} A_z] \vec{e}_{\vartheta}
$$
$$
\qquad \qquad + [\frac{1}{\rho} \partial_{\rho}(\rho A_{\vartheta}) -
   \frac{1}{\rho} \partial_{\vartheta} A_{\rho}] \vec{e}_z .
$$
The vector of Pauli matrices is
$$
\hat{\vec{\sigma}}
=
\vec{e}_{\rho} \, \hat{\vec{\sigma}}_{\rho} +
\vec{e}_{\vartheta} \, \hat{\vec{\sigma}}_{\vartheta} +
\vec{e}_z \, \hat{\vec{\sigma}}_z,
$$
where
$$
\hat{\sigma}_{\rho} = \left(\begin{array}{cc} 0 & e^{-i\vartheta}
\\ e^{i\vartheta}  & 0 \end{array}
  \right)
  \quad,
\hat{\sigma}_{\vartheta} = i\,\left(\begin{array}{cc} 0 &  -e^{-i\vartheta}
\\ e^{i\vartheta}  & 0 \end{array}
  \right)
  \quad,
\hat{\sigma}_z = \left(\begin{array}{cc} 1 & 0 \\ 0 &-1 \end{array}
  \right)
$$

\subsection{Single-particle wave function in cylindrical coordinates}
\label{sec:wf_cyl}

The single-particle particle wave function and its time reversal
are expressed in the cylindrical coordinates as spinors
\begin{equation}
\Psi_{i}(\vec r) =
\left(\begin{array}{c} {R_{i}^{(+)}(\rho,z) e^{im_{i}^{(+)}\vartheta}} \\
{R_{i}^{(-)}(\rho,z) e^{im_{i}^{(-)}\vartheta}} \end{array}
 \right) ,
\end{equation}

\begin{equation}
\Psi_{\overline{i}}(\vec r) = \hat T \Psi_{i}(\vec r) =
\left(\begin{array}{c}
{-R_{i}^{(-)}(\rho,z) e^{-im_{i}^{(-)}\vartheta}} \\
{R_{i}^{(+)}(\rho,z) e^{-im_{i}^{(+)}\vartheta}} \end{array}
 \right)
\end{equation}
where $K_i$ is the projection of the complete single-particle
moment onto symmetry z-axis of the axial nucleus.

In short notations covering both the state $i$ and time-inverse
$\bar{i}$ states, the spinors read
\begin{equation}\label{eq:tilde_R}
\tilde{R}_{i}^{(\sigma )}(\rho , z)
e^{i\tilde{m}_{i}^{(\sigma )}\vartheta} = \left(
\begin{array}{c}
{R_{i}^{(\sigma )}(\rho,z) e^{im_{i}^{(\sigma )}\vartheta}}
\quad \mbox{for ordinary state} \quad i
\nonumber \\
{-\sigma R_{i}^{(-\sigma )}(\rho,z) e^{-im_{\bar{i}}^{(-\sigma )}
\vartheta}} \quad \mbox{for time reversal state} \quad \bar{i}
\end{array}
\right)
\end{equation}
where
\begin{equation}
m_{i}^{(\sigma)} = K - \frac{1}{2} \sigma ,
\quad
m_{i}^{(-\sigma)} = m_{i}^{(\sigma)} + \sigma ,
\quad
m_{\bar i}^{(\sigma)} = -m_{i}^{(-\sigma)} .
\label{eq:new_m}
\end{equation}

\section{Useful relations} 

\subsection{Hermitian and time-conjugate properties}
\label{sec:herm+time-conj}

All the operators used in the model are hermitian
$(\hat A = \hat A^{\dagger})$
and  have the definite time density
$$
\hat{\bar A} = T \hat A T^{-1} =\gamma_T^A \hat A, \quad \gamma_T^A=\pm 1 \;
$$
where $T$ is the time-inversion operator: $T|i>=|\bar i>, \quad  T|\bar i>=-|i>$.

There are useful transmutation relation for
the hermitian operators
\begin{equation}\label{eq:herm}
<j|\hat{A}^{\dagger}|i>=<j|\hat{A}|i>=<i|\hat{A}|j>^* \;
\end{equation}
and relations for time-inverse states and operators
\begin{eqnarray}\label{eq:time_rev}
<{\bar j}|{\bar A}|{\bar i}>&=&<{\bar j}|T^{-1} T{\bar A}T^{-1} T|{\bar i}>
=<T{\bar j}|T{\bar A}T^{-1} T|{\bar i}>^*=<j|A|i>^* ,
\nonumber\\
<{\bar j}|{\bar A}|i>&=& - <j|A|{\bar i}>^* \; .
\end{eqnarray}
In axial even-even nuclei, the densities are built
from the pairs
\begin{equation}\label{eq:diag}
<i|A|i> + <{\bar i}|A|{\bar i}> = <i|A|i> + \gamma_T^A<i|A|i>^*
= <i|A|i>(1 + \gamma_T^A) \;
\end{equation}
and hence vanish for time-odd operators.

\subsection{Connections for operators with definite time-parity}
\label{sec:T_par_oper}

 Operators with the definite time-parity have the useful property
\begin{equation}\label{eq:<[A,B]>=0_2}
  <0|[\hat{A},\hat{B}]|0> = <0|\hat{A}\hat{B}|0> (1-\gamma_T^A\gamma_T^B)
\end{equation}
i.e. the average value of the commutator survives only if operators $\hat{A}$
and $\hat{B}$ are of the opposite time-parity. Indeed,
\begin{eqnarray}
&&<0|[\hat{A},\hat{B}]|0> = <0|\hat{A}\hat{B}|0> - <0|\hat{B}\hat{A}|0>
\nonumber\\
&=& <0|\hat{A}\hat{B}|0> -<0|T^{-1} T\hat{B}T^{-1} T\hat{A}T^{-1} T|0>
= <0|\hat{A}\hat{B}|0> -\gamma^A_T \gamma^B_T <0|T^{-1}\hat{B}\hat{A}T|0>
\nonumber\\
&=& <0|\hat{A}\hat{B}|0> - \gamma^A_T \gamma^B_T <T0|\hat{B}\hat{A}|T0>^*
= <0|\hat{A}\hat{B}|0> - \gamma^A_T \gamma^B_T <0|\hat{B}\hat{A}|0>^*
\nonumber\\
&=& <0|\hat{A}\hat{B}|0> - \gamma^A_T \gamma^B_T <0|(\hat{B}\hat{A})^+|0>
= <0|\hat{A}\hat{B}|0>  (1-\gamma^A_T \gamma^B_T) \; .
\end{eqnarray}

\subsection{Representation through matrix elements}
\label{sec:rep_me}

Responses
(\ref{eq:X})-(\ref{eq:Y})
and inverse strength matrices (\ref{eq:kappa})-(\ref{eq:eta})
read as the averaged commutators
\begin{equation}\label{eq:<[A,B]>}
  \langle 0|[\hat{A},\hat{B}]|0\rangle  \quad \mbox{with} \quad
 \gamma_T^A= - \gamma_T^B  .
\end{equation}
Calculation  of these values can be greatly simplified if to express them
through the matrix elements of the operators
$\hat{A}$ and $\hat{B}$, because in practice these matrix elements are already
in our disposal.
In this case,  the commutator reads
\begin{eqnarray}\label{eq:AB_cc}
   <0|[\hat{A}_s,\hat{B}_s]|0> &=&  \sum_{ij}
\{
<0|\hat{A}_s|ij><ij|\hat{B}_s|0> -
<0|\hat{B}_s|ij><ij|\hat{A}_s|0>
\}
\\
&=&  2 \sum_{ij}^{K_i,K_j>0}
\{
<ij|\hat{A}_s|0>^*<ij|\hat{B}_s|0> - <ij|\hat{A}_s|0><ij|\hat{B}_s|0>^*
\} .
\nonumber
\end{eqnarray}
The sum runs all the states, both ordinary and time reversed.
Using the properties (\ref{eq:herm})-(\ref{eq:time_rev}), the matrix
elements $<\bar i\bar j|\hat{A}_s|0>$
and $<\bar i j|\hat{A}_s|0>$ are reduced to $<ij|\hat{A}_s|0>$ and
$<i\bar j|\hat{A}_s|0>$, respectively. Hence the coefficient 2.

In the SRPA, the operator $\hat A$ is associated with operators
${\hat Q}$ or ${\hat P}$, which have real or imagine matrix elements,
respectively. Then
\begin{eqnarray}\label{eq:AB_T_ev}
<0|[\hat{Q}_{sk},\hat{B}_s]|0> &=&
2 \sum_{ij}^{K_i,K_j>0}
 <ij|\hat{Q}_{sk}|0> (<ij|\hat{B}_s|0> - <ij|\hat{B}_s|0>^*)
\nonumber\\
&=& 4i  \sum_{ij}^{K_i,K_j>0}
 <ij|\hat{Q}_{sk}|0> \; \Im\{<ij|\hat{B}_s|0>\} \; ,
\end{eqnarray}
\begin{eqnarray}\label{eq:AB_T_odd}
<0|[\hat{P}_{sk},\hat{B}_s]|0> &=&
- 2 \sum_{ij}^{K_i,K_j>0}
 <ij|\hat{P}_{sk}|0> (<ij|\hat{B}_s|0> + <ij|\hat{B}_s|0>^*)
\nonumber\\
&=& -4  \sum_{ij}^{K_i,K_j>0}
 <ij|\hat{P}_{sk}|0> \; \Re\{<ij|\hat{B}_s|0>\} \;
\end{eqnarray}
where $\Im\{ ... \}$ and $\Re\{ ... \}$ mean the imagine and real parts of the values in the
parenthesis.

Then the strength matrices read
\begin{eqnarray}
\label{eq:kappa_me}
\kappa_{s'k',sk}^{-1 }&=&
- i <0|[\hat{P}_{s'k'},{\hat X}_{sk}^{s'}]|0>
= 4i \sum_{ij}^{K_i , K_j >0}
<ij|\hat{P}_{s'k'}|0> \;
\Re\{<ij|\hat{X}_{sk}^{s'}]|0>\}
\nonumber \\
&=& -4 \sum_{ij}^{K_i , K_j >0}
\overline{<ij|\hat{P}_{s'k'}|0>} \;
\overline{<ij|\hat{X}_{sk}^{s'}]|0>}
\; ,
\\
\label{eq:eta_me}
\eta_{s'k',sk}^{-1 }&=&
-i \langle 0|[\hat{Q}_{k'},{\hat Y}_{k}]|0>
= 4 \sum_{ij}^{K_i , K_j >0}
<ij|\hat{Q}_{s'k'}|0> \;
\Im\{<ij|\hat{Y}_{sk}^{s'}]|0>\}
\nonumber \\
&=&
4 \sum_{ij}^{K_i , K_j >0}
\overline{<ij|\hat{Q}_{s'k'}|0>} \;
\overline{<ij|\hat{Y}_{sk}^{s'}]|0>} \;
\end{eqnarray}
where the overline matrix elements
\begin{eqnarray}\label{eq:over_P,X}
\overline{<ij|\hat{P}_{s'k'}|0>}&=& -i<ij|\hat{P}_{s'k'}|0>,
\quad
\overline{<ij|\hat{X}_{sk}^{s'}]|0>} = <ij|\hat{X}_{sk}^{s'}]|0>,
\\
\overline{<ij|\hat{Q}_{s'k'}|0>} &=& <ij|\hat{Q}_{s'k'}|0>,
\qquad
\overline{<ij|\hat{Y}_{sk}^{s'}]|0>} = -i <ij|\hat{Y}_{sk}^{s'}]|0>
\label{eq:over_Q,Y}
\end{eqnarray}
are real. It is seen that both strength matrices (\ref{eq:kappa_me}) and
(\ref{eq:eta_me}) are real as well.

The case of responses is more involved in the sense that matrix elements of
the second operator in the commutator are transition densities
(\ref{trans_dens_j})-(\ref{trans_dens_Im}) which are generally
complex. The responses entering $\hat X$ and  $\hat Y$ operators read
\begin{eqnarray}\label{eq:R_T_ev}
{\cal R}^{\alpha}_{X,sk} &=& i <0|[\hat{P}_{sk},\hat{J}^{\alpha}_s]|0>  =
-4i \sum_{ij}^{K_i,K_j >0}
<ij|\hat{P}_{sk}|0> \; \Re\{ <ij|\hat{J}^{\alpha}_s|0> \}
\nonumber\\
&=& 4  \sum_{ij}^{K_i,K_j >0}
\overline{<ij|\hat{P}_{sk}|0>} \; \Re\{ <ij|\hat{J}^{\alpha}_s|0> \} ,
\\
\label{eq:R_T_odd}
{\cal R}^{\alpha}_{Y,sk} &=& i <0|[\hat{Q}_{sk},\hat{J}^{\alpha}_s]|0>
= -4  \sum_{ij}^{K_i,K_j >0}
<ij|\hat{Q}_{sk}|0> \; \Im\{ <ij|\hat{J}_{\alpha}|0> \}
\nonumber\\
&=& -4  \sum_{ij}^{K_i,K_j >0}
\overline{<ij|\hat{Q}_{sk}|0>} \; \Im\{ <ij|\hat{J}^{\alpha}_s|0> \}
\end{eqnarray}
where $<ij|\hat{J}^{\alpha}_s|0>$ are transition densities.
It is seen that all the responses are real.

Following (\ref{trans_dens_j})-(\ref{trans_dens_Im}), the transition densities
(or their components) for $\mu= K_i \pm K_j $ are proportional to
$e^{-i\mu\vartheta} = \cos \mu\vartheta - i \sin \mu\vartheta$
or
$i e^{-i\mu\vartheta} = i\cos \mu\vartheta + \sin \mu\vartheta$.
Hence  the response obviously vanishes at $\mu =0$,
if its $\Im\{ ... \}$ or $\Re\{ ... \}$
part delivers $\sin \mu\vartheta$.

\section{Densities and currents for Skyrme functional}
\label{sec:dens_curr}

In Skyrme forces, the complete set of the densities involves the ordinary
density, kinetic-energy density, spin-orbital density, current density and  spin
density:
\begin{eqnarray*}
  \rho_s({\vec r},t)
  &=&
  \sum_{h \epsilon s}^{occ}
  \varphi^*_h({\vec r},t)\varphi_h^{\mbox{}}({\vec r},t) ,
  \qquad\qquad\qquad\qquad\qquad\qquad\quad
  \hat{T} \rho \hat{T}^{-1}=\rho
\\
  \tau_s({\vec r},t)
  &=&
  \sum_{h \epsilon s}^{occ}
  \vec{\nabla}\varphi^*_h({\vec r},t)\!\cdot\!
  \vec{\nabla}\varphi_h^{\mbox{}}({\vec r},t) ,
 \qquad\qquad\qquad\qquad\qquad\quad
  \hat{T} \tau \hat{T}^{-1}=\tau
\\
  \vec{\Im}_s({\vec r},t)
  &=&
  -i\sum_{h \epsilon s}^{occ}
  \varphi^*_h({\vec r},t)(\vec{\nabla}\times
  \hat{\vec{\sigma}})\varphi_h^{\mbox{}}({\vec r},t) ,
  \qquad\quad\qquad\qquad\quad
  \hat{T} {\vec \Im} \hat{T}^{-1}={\vec \Im}
\\
  \vec{j}_s({\vec r},t)
  &=&
  \frac{-i}{2}\sum_{h \epsilon s}^{occ}
  \left[
  \varphi^*_h({\vec r},t)\vec{\nabla}\varphi_h^{\mbox{}}({\vec r},t)
  \!-\!
  \vec{\nabla}\varphi^*_h({\vec r},t)\varphi_h^{\mbox{}}({\vec r},t)
  \right] ,
\quad
  \hat{T} {\vec j} \hat{T}^{-1}=-{\vec j}
\\
  \vec{\sigma}_s({\vec r})
  &=&
  \sum_{h \epsilon s}^{occ}
  \varphi^*_h({\vec r},t)\hat{\vec{\sigma}}
  \varphi_h^{\mbox{}}({\vec r},t) ,
\qquad\qquad\qquad\qquad\qquad\quad \;\;
  \hat{T} {\vec \sigma} \hat{T}^{-1}=-{\vec \sigma}
\end{eqnarray*}
where the sum runs over the occupied (hole) single-particle states $h$.
The  associated {\it hermitian} operators are
\begin{eqnarray*}
  \hat{\rho}_s(\vec{r})
  &=&
  \sum_{i=1}^{N_s}\delta(\vec{r}_i-\vec{r}) ,
\\
  \hat{\tau}_s(\vec{r})
  &=&
  \sum_{i=1}^{N_s}
  \overleftarrow{\nabla}\delta(\vec{r}_i - \vec{r})\vec{\nabla} ,
\\
  \hat{\vec{\Im}}_s(\vec{r})
  &=&
  \sum_{i=1}^{N_s}
  \delta(\vec{r_i} - \vec{r})\vec{\nabla}\!\times\!\hat{\vec{\sigma}} ,
\\
  \hat{\vec{j}}_s(\vec{r})
  &=&
  \frac{1}{2}\sum_{i=1}^{N_s}
  \left\{ \vec{\nabla},
  \delta(\vec{r}_i-\vec{r})
  \right\} ,
\\
  \hat{\vec{\sigma}}_s(\vec{r})
  &=&
  \sum_{i=1}^{N_s} \delta(\vec{r}_i-\vec{r})\hat{\vec{\sigma}} \; .
\end{eqnarray*}
where $\hat{\vec{\sigma}}$ is the Pauli matrix, $N_s$ is number of protons or neutrons
in the nucleus.

\section{Transition densities}
\label{sec:trans_dens_pair}

\subsection{Explicit expressions}

For $\mu=|K_i-K_j|$ with  $K_i, K_j > 0$,
the transition densities read
\begin{eqnarray}\label{trans_dens_j}
<ij|\hat{\vec{j}}_s|\tilde 0> (\vec r)&=&
 [\vec{e}_{\rho} \cdot i j^s_{ij;\,\,\rho}(\rho,z)
+  \vec{e}_{z} \cdot i j^s_{ij;\,\,z}(\rho,z)
+  \vec{e}_{\vartheta} \cdot j^s_{ij;\,\,\vartheta}(\rho,\,\,z)]
e^{-i(K_i-K_j)\vartheta} \; ,\quad
\\
\label{trans_dens_s}
<ij|\hat{\vec{s}}_s|\tilde 0>(\vec r)&=&
 [\vec{e}_{\rho} \cdot s^s_{ij;\,\,\rho}(\rho,z)
+  \vec{e}_{z} \cdot s^s_{ij;\,\,z}(\rho,z)
+  \vec{e}_{\vartheta} \cdot i s^s_{ij;\,\,\vartheta}(\rho,\,\,z)]
e^{-i(K_i-K_j)\vartheta} \; ,\quad
\\
\label{trans_dens_rho}
<ij|\hat{\rho}_s|\tilde 0>(\vec r)&=& \rho^s_{ij}(\rho,z)
e^{-i(K_i-K_j)\vartheta} \; ,\quad
\\
\label{trans_dens_tau}
<ij|\hat{\tau}_s|\tilde 0>(\vec r)&=& \tau^s_{ij}(\rho,z)
e^{-i(K_i-K_j)\vartheta} \; ,\quad
\\
\label{trans_dens_Im}
<ij|\hat{\vec{\Im}}_s|\tilde 0>(\vec r)&=&
 [\vec{e}_{\rho} \cdot \Im^s_{ij;\,\,\rho}(\rho,z)
+  \vec{e}_{z} \cdot \Im^s_{ij;\,\,z}(\rho,z)
+  \vec{e}_{\vartheta} \cdot i \Im^s_{ij;\,\,\vartheta}(\rho,\,\,z)]
e^{-i(K_i-K_j)\vartheta}
\end{eqnarray}
where the components with low indices $ij$ are real.
The case of
the time-inverse state $\bar j$ is straightforwardly obtained
by $K_j \to -K_j$. As is shown below,
the components accompanied by the imagine unit
($j^s_{ij;\,\,\rho}, j^s_{ij;\,\,z}, s^s_{ij;\,\,\theta}$,
and $\Im^s_{ij;\,\,\theta}$)
demonstrate the specific features.

Using wave functions from the Appendix \ref{sec:wf_cyl} and density
operators from the Appendix \ref{sec:dens_curr}, one gets expressions for the
real components of the transition densities (\ref{trans_dens_j})-(\ref{trans_dens_Im})
and for their transmutation properties:
\begin{eqnarray}\label{td_j_rho_pair}
j^s_{ij;\,\,\rho}(\rho,z) &&=\, \frac{1}{2} \;
{\cal U}^{(-)}_{ij}
 \sum_{\sigma =+,-}
\Bigl[\, (\partial_{\rho} \tilde{R}_i^{(\sigma )})
\tilde{R}_j^{(\sigma )} \,-
 \tilde{R}_i^{(\sigma )}(\partial_{\rho} \tilde{R}_j^{(\sigma )})
\,\Bigr] ,
 \\
\label{td_j_z_pair}
j^s_{ij;\,\,z}(\rho,\,\,z) &&=\, \frac{1}{2} \;
{\cal U}^{(-)}_{ij}
 \sum_{\sigma =+,-}
\Bigl[\, (\partial_{z}  \tilde{R}_i^{(\sigma )})
\tilde{R}_j^{(\sigma )} \,-
  \tilde{R}_i^{(\sigma )} (\partial_{z}  \tilde{R}_j^{(\sigma )})
\,\Bigr] ,
\\
\label{td_j_theta_pair}
j^s_{ij;\,\,\vartheta}(\rho,\,\,z) &&=\, \frac{1}{2} \;
{\cal U}^{(-)}_{ij}
\sum_{\sigma =+,-} \Bigl[\, \frac{ \tilde{R}_i^{(\sigma )}
\tilde{R}_j^{(\sigma )}}{\rho} (\tilde{m}_i^{(\sigma)} + \tilde{m}_j^{(\sigma)})
\,\Bigr] .
\end{eqnarray}
$$
\begin{array}{ccccccc}
j^s_{ji;\,\,\rho}(\rho,z)&=& j^s_{ij;\,\,\rho}(\rho,z), \quad
j^s_{ji;\,\,z}(\rho,\,\,z)= j^s_{ij;\,\,z}(\rho,\,\,z), \quad
j^s_{ji;\,\,\vartheta}(\rho,\,\,z)= -j^s_{ij;\,\,\vartheta}(\rho,\,\,z).
\end{array}
$$
\begin{eqnarray}
&&s^s_{ij;\,\,\rho}(\rho,z) \,=\,
{\cal U}^{(-)}_{ij}
\sum_{\sigma =+,-} \Bigl[\,\tilde{R}_i^{(-\sigma )}
\tilde{R}_j^{(\sigma )} \,\Bigr] ,
\label{td_s_rho_pair}\\
&&s^s_{ij;\,\,z}(\rho,\,\,z) \,=\,
{\cal U}^{(-)}_{ij}
\sum_{\sigma =+,-}\sigma \Bigl[\,\tilde{R}_i^{(\sigma )}
\tilde{R}_j^{(\sigma )}  \,\Bigr] ,
\label{td_s_z_pair} \\
&&s^s_{ij;\,\,\vartheta}(\rho,\,\,z) \,=\,
{\cal U}^{(-)}_{ij}
\sum_{\sigma =+,-} \sigma \Bigl[\,\tilde{R}_i^{(-\sigma )}
\tilde{R}_j^{(\sigma )}\,\Bigr] .
\label{td_s_theta_pair}
\end{eqnarray}
$$
\begin{array}{ccccccc}
s^s_{ji;\,\,\rho}(\rho,z)&=& - s^s_{ij;\,\,\rho}(\rho,z), \quad
s^s_{ji;\,\,z}(\rho,\,\,z)= - s^s_{ij;\,\,z}(\rho,\,\,z), \quad
s^s_{ji;\,\,\vartheta}(\rho,\,\,z)= s^s_{ij;\,\,\vartheta}(\rho,\,\,z) .
\end{array}
$$
\begin{eqnarray}\label{td_rho_pair}
\rho^s_{ij}(\rho,z) \,=\,
{\cal U}^{(+)}_{ij}
 \sum_{\sigma =+,-}
\Bigl[\,\tilde{R}_i^{(\sigma)}\tilde{R}_j^{(\sigma)}\,\Bigr] .
\end{eqnarray}
$$
\rho^s_{ ji}(\rho,z) = \rho^s_{ ij}(\rho,z) .
$$
\begin{eqnarray}\label{td_tau_pair}
\tau^s_{ij}(\rho,z) \,=\,
{\cal U}^{(+)}_{ij}
 \sum_{\sigma =+,-}
\Bigl[
(\partial_{\rho}\tilde{R}_i^{(\sigma)})(\partial_{\rho}\tilde{R}_j^{(\sigma)})+
(\partial_{z}\tilde{R}_i^{(\sigma)})
(\partial_{z}\tilde{R}_j^{(\sigma)})
+\frac{\tilde{m}_i^{(\sigma)}\tilde{m}_j^{(\sigma)}}{\rho^2}
\tilde{R}_i^{(\sigma)}\tilde{R}_j^{(\sigma)}
\Bigr] .
\end{eqnarray}
$$
\tau^s_{ ji}(\rho,z)=  \tau^s_{ ij}(\rho,z) .
$$
\begin{eqnarray}\label{td_I_rho_pair}
&&\Im^s_{ij; \rho}(\rho,z) \,=\, \frac{1}{2} \;
{\cal U}^{(+)}_{ij}
\sum_{\sigma =+,-} \sigma \Bigl[\,
\tilde{R}_i^{(\sigma)} (\partial_{z} \tilde{R}_j^{(-\sigma)})
+
\tilde{R}_j^{(\sigma)}(\partial_{z} \tilde{R}_i^{(-\sigma
)})
\nonumber \\
&&\qquad \qquad \qquad \qquad \qquad \qquad
+ (\tilde{m}_i^{(\sigma)} + \tilde{m}_j^{(\sigma)})
\frac{\tilde{R}_i^{(\sigma)} \tilde{R}_j^{(\sigma)}}{\rho}
 \Bigr] ,
\\
\label{td_I_z_pair}
&&\Im^s_{ij; z}(\rho,z) \,=\, \frac{1}{2} \;
{\cal U}^{(+)}_{ij}
\sum_{\sigma =+,-}  \Bigl[\,
  \sigma
\Bigl(
\tilde{R}_i^{(-\sigma)} (\partial_{\rho} \tilde{R}_j^{(\sigma )})
+  \tilde{R}_j^{(-\sigma)}(\partial_{\rho} \tilde{R}_i^{(\sigma )})
 \Bigr)
\nonumber \\
&&\qquad \qquad \qquad \qquad \qquad \qquad
-  (\tilde{m}_i^{(-\sigma)} + \tilde{m}_j^{(\sigma)})
\frac{\tilde{R}_i^{(\sigma)}\tilde{R}_j^{(-\sigma)}}{\rho}
\Bigr] ,
\\
\label{td_I_theta_pair}
&& \Im^s_{ij;\,\,\vartheta}(\rho,z) \,=\, \frac{1}{2} \;
{\cal U}^{(+)}_{ij}
\sum_{\sigma =+,-} \Bigl[\, \sigma
\Bigl(
\tilde{R}_i^{(\sigma )}(\partial_{\rho} \tilde{R}_j^{(\sigma )})
+
\tilde{R}_j^{(-\sigma )}(\partial_{\rho} \tilde{R}_i^{(-\sigma )})
\Bigr)
\nonumber \\
&&\qquad \qquad \qquad \qquad \qquad \qquad
- \tilde{R}_i^{(\sigma)}(\partial_{z} \tilde{R}_j^{(-\sigma )})
+ \tilde{R}_j^{(-\sigma )} (\partial_{z} \tilde{R}_i^{(\sigma )})
\Bigr] .
\end{eqnarray}

$$
\begin{array}{ccccccc}
\Im^s_{ji;\,\,\rho}(\rho,z)&=& \Im^s_{ij;\,\,\rho}(\rho,z), \quad
\Im^s_{ji;\,\,z}(\rho,\,\,z)&=& \Im^s_{ij;\,\,z}(\rho,\,\,z), \quad
\Im^s_{ji;\,\,\vartheta}(\rho,\,\,z)&=&-
\Im^s_{ij;\,\,\vartheta}(\rho,\,\,z) .
\end{array}
$$

\subsection{Useful features}

Some components of the vector responses become zero at $\mu =0$.
These cases can be revealed by inspecting the specific combinations
of the transition densities involved into responses.
Following (\ref{eq:R_T_ev})-(\ref{eq:R_T_odd}), these
combinations are
$$
<ij|\hat{J}^{\alpha}_s|\tilde 0>+<ij|\hat{J}^{\alpha}_s|\tilde 0>^*= 2  \;
\Re\{ <ij|\hat{J}^{\alpha}_s|\tilde 0> \}
$$
for time-even densities
$$
<ij|\hat{J}^{\alpha}_s|\tilde 0>-<ij|\hat{J}^{\alpha}_s|\tilde 0>^*= 2i \;
\Im \{ <ij|\hat{J}^{\alpha}_s|\tilde 0> \}
$$
and for time-odd ones.

Specifically, these combinations read
\begin{eqnarray}
&& <ij|\hat{\vec{j}}_s|\tilde 0>-<ij|\hat{\vec{j}}_s|\tilde 0>^*
\nonumber\\
&=&
2i[
(\vec{e}_{\rho} \cdot j^s_{ij;\,\,\rho}(\rho,z)
+  \vec{e}_{z} \cdot j^s_{ij;\,\,z}(\rho,z))
\cos \mu\vartheta
-
 \vec{e}_{\vartheta} \cdot j^s_{ij;\,\,\vartheta}(\rho,\,\,z)
\sin \mu\vartheta
] ,
\nonumber\\
&& <ij|\hat{\vec{s}}_s|\tilde 0>-<ij|\hat{\vec{s}}_s|\tilde 0>^*
\nonumber\\
&=&
2i [ - (\vec{e}_{\rho} \cdot s^s_{ij;\,\,\rho}(\rho,z)
+  \vec{e}_{z} \cdot s^s_{ij;\,\,z}(\rho,z))
\sin \mu\vartheta
+
 \vec{e}_{\vartheta} \cdot s^s_{ij;\,\,\vartheta}(\rho,\,\,z)
\cos \mu\vartheta
] .
\nonumber\\
&& <ij|\hat{\rho}_s|\tilde 0>+<ij|\hat{\rho}_s|\tilde 0>^*
= 2 \rho^s_{ij}(\rho,z)\cos \mu\vartheta ,
\nonumber\\
&& <ij|\hat{\tau}_s|\tilde 0>+<ij|\hat{\tau}_s|\tilde 0>^*
= 2 \tau^s_{ij}(\rho,z)\cos \mu\vartheta  ,
\nonumber\\
&& <ij|\hat{\vec{\Im}}_s|\tilde 0>+<ij|\hat{\vec{\Im}}_s|\tilde 0>^*
\nonumber\\
&=&
2 [ (\vec{e}_{\rho} \cdot\Im^s_{ij;\,\,\rho}(\rho,z)
+  \vec{e}_{z} \cdot \Im^s_{ij;\,\,z}(\rho,z))
\cos \mu\vartheta
-
 \vec{e}_{\vartheta} \cdot\Im^s_{ij;\,\,\vartheta}(\rho,\,\,z)
\sin \mu\vartheta
]  \; .
\nonumber
\end{eqnarray}
For $\mu =0$, the combinations with $\sin \mu\vartheta $ vanish
\begin{eqnarray}\label{eq:comb_Im_j=0}
(<ij|\hat{\vec{\Im}}_s|\tilde 0>+<ij|\hat{\vec{\Im}}_s|\tilde 0>^*)_{\theta} &=& 0 ,
\quad
(<ij|\hat{\vec{j}}_s|\tilde 0>-<ij|\hat{\vec{j}}_s|\tilde 0>^*)_{\theta} = 0 ,
\\
(<ij|\hat{\vec{s}}_s|\tilde 0>-<ij|\hat{\vec{s}}_s|\tilde 0>^*)_{\rho} &=& 0 ,
\quad
(<ij|\hat{\vec{s}}_s|\tilde 0>-<ij|\hat{\vec{s}}_s|\tilde 0>^*)_{z} = 0
\label{eq:comb_s=0}
\end{eqnarray}
and hence the corresponding response components
\\
\\
\fbox{
\parbox{16cm}{
\begin{equation}\label{eq:res=0_mu=0}
 \mu=0 \quad \rightarrow \quad
j_{Y, sk}^{\theta} = s_{Y, sk}^{z} = s_{Y, sk}^{\rho} =
\Im_{Y, sk}^{\theta}  = 0.
\end{equation}
}}
\\

The curl in cylindrical coordinates is
\begin{equation}
rot \vec{A} =  \nabla \times \vec{A} =
[ \frac{1}{\rho} \partial_{\vartheta} A_z
   - \partial_z A_{\vartheta}] \vec{e}_{\rho} +
[ \partial_z A_{\rho} - \partial_{\rho} A_z] \vec{e}_{\vartheta}
+ [\frac{1}{\rho} \partial_{\rho}(\rho A_{\vartheta}) -
   \frac{1}{\rho} \partial_{\vartheta} A_{\rho}] \vec{e}_z .
\end{equation}
Then, taking into account (\ref{eq:comb_Im_j=0})-(\ref{eq:comb_s=0})
and zero value of the derivatives  $\partial_{\theta}$ from all the
transition densities at $\mu =0$, one gets
\\
\\
\fbox{
\parbox{16cm}{
\begin{eqnarray} \label{eq:rot_current}
 \mu = 0 \quad \to \quad
&&\{ \nabla \times
(<ij|{\vec j}_s|\tilde 0> - <ij|{\vec j}_{s}|\tilde 0>^*)
\}_{\rho} = 0 ,
\\
&&\{ \nabla \times
 (<ij|{\vec j}_s|\tilde 0> - <ij|{\vec j}_s|\tilde 0>^*)
\}_{z} = 0 ,
\nonumber\\
&&\{ \nabla \times
(<ij|{\vec s}_{s}|\tilde 0> - <ij|{\vec s}_{s}|\tilde 0>^*)
\}_{\theta} = 0 ,
\nonumber\\
&&\{ \nabla \times
(<ij|{\vec \Im}_{s}|\tilde 0> + <ij|{\vec \Im}_{s}|\tilde 0>^*)
\}_{\rho} = 0 ,
\nonumber\\
&&\{ \nabla \times
(<ij|{\vec \Im}_{s}|\tilde 0> + <ij|{\vec \Im}_{s}|\tilde 0>^*)
\}_{z} = 0 .
\nonumber
\end{eqnarray}
}}
\\
\\
These relations are used in derivation of the first and second functional derivatives
of the Skyrme functional (terms with $b_4$ and $b_4'$).

\section{Responses}

The responses have the general form (\ref{eq:R_T_ev})-(\ref{eq:R_T_odd}).
The explicit expressions read
\begin{eqnarray}
\vec{j}_{Y, sk}(\vec{r})
&=&
  \vec{e}_{\rho}j_{Y, sk}^{\rho}(\vec{r})
+ \vec{e}_{z}j_{Y, sk}^{z}(\vec{r})
+ \vec{e}_{\theta}j_{Y, sk}^{\theta}(\vec{r})
\nonumber\\
&=&
  \vec{e}_{\rho}j_{Y, sk}^{\rho}(\rho,z)\cos\mu\theta
+ \vec{e}_{z}j_{Y, sk}^{z}(\rho,z)\cos\mu\theta
+ \vec{e}_{\theta}j_{Y, sk}^{\theta}(\rho,z)\sin\mu\theta \; ,
\label{vec_j_Y}
\\
\vec{s}_{Y, sk}(\vec{r})
&=&
  \vec{e}_{\rho}s_{Y, sk}^{\rho}(\vec{r})
+ \vec{e}_{z}s_{Y, sk}^{z}(\vec{r})
+ \vec{e}_{\theta}s_{Y, sk}^{\theta}(\vec{r})
\nonumber\\
&=&
  \vec{e}_{\rho}s_{Y, sk}^{\rho}(\rho,z)\sin\mu\theta
+ \vec{e}_{z}s_{Y, sk}^{z}(\rho,z)\sin\mu\theta
+ \vec{e}_{\theta}s_{Y, sk}^{\theta}(\rho,z)\cos\mu\theta \; ,
\label{vec_s_Y}
\nonumber\\
\rho_{X, sk}(\vec{r})
&=& \rho_{X, sk}(\rho,z)\cos\mu\theta
\nonumber\\
\tau_{X, sk}(\vec{r})
&=& \tau_{X, sk}(\rho,z)\cos\mu\theta
\nonumber\\
\vec{\Im}_{X, sk}(\vec{r})
&=&
  \vec{e}_{\rho}\Im_{X, sk}^{\rho}(\vec{r})
+ \vec{e}_{z}\Im_{X, sk}^{z}(\vec{r})
+ \vec{e}_{\theta}\Im_{X, sk}^{\theta}(\vec{r})
\nonumber\\
&=&
  \vec{e}_{\rho}\Im_{X, sk}^{\rho}(\rho,z)\cos\mu\theta
+ \vec{e}_{z}\Im_{X, sk}^{z}(\rho,z)\cos\mu\theta
+ \vec{e}_{\theta}\Im_{X, sk}^{\theta}(\rho,z)\sin\mu\theta \; ,
\label{vec_J_Y}
\end{eqnarray}
The components of time-odd responses have the form
\begin{eqnarray}
j_{Y, sk}^{\rho}(\vec r)&=& i <\tilde{0}|[\hat{Q}_{sk},{\hat j}_s^{\rho}]|\tilde{0}> =
-4 \sum_{ij}^{K_i , K_j>0}
<ij|\hat{Q}_{sk}|\tilde 0>   \; \Im\{<ij|\hat{j}_s^{\rho}]|\tilde 0>\}
\nonumber\\
&=& -4 \sum_{ij}^{K_i , K_j >0}
\overline{<ij|\hat{Q}_{sk}|\tilde 0>} \;
j^s_{ij;\,\,\rho}(\rho,z) \cos \mu\theta
= j_{Y, sk}^{\rho}(\rho,z)\cos \mu\theta ,
\label{eq:resp_j_rho_qp}\\
j_{Y, sk}^{z}(\vec r)&=& i <\tilde{0}|[\hat{Q}_{sk},{\hat j}_s^{z}]|\tilde{0}> =
-4 \sum_{ij}^{K_i , K_j>0}
<ij|\hat{Q}_{sk}|\tilde 0>  \; \Im\{<ij|\hat{j}_s^{z}]|\tilde 0>\}
\nonumber\\
&=& -4 \sum_{ij}^{K_i , K_j >0}
\overline{<ij|\hat{Q}_{sk}|\tilde 0>} \;
j^s_{ij;\,\,z}(\rho,z) \cos \mu\theta
= j_{Y_, sk}^{z}(\rho,z)\cos \mu\theta ,
\label{eq:resp_j_z_qp}\\
j_{Y, sk}^{\theta}(\vec r)&=& i <\tilde{0}|[\hat{Q}_{sk},{\hat j}_s^{\theta}]|\tilde{0}> =
-4 \sum_{ij}^{K_i , K_j>0}
<ij|\hat{Q}_{sk}|\tilde 0>   \; \Im\{<ij|\hat{j}_s^{\theta}]|\tilde 0>\}
\nonumber\\
&=& 4 \sum_{ij}^{K_i , K_j >0}
\overline{<ij|\hat{Q}_{sk}|\tilde 0>} \;
j^s_{ij;\,\,\theta}(\rho,z)  \sin \mu\theta
= j_{Y, sk}^{\theta}(\rho,z) \sin \mu\theta ,
\label{eq:resp_j_theta_qp}\\
s_{Y, sk}^{\rho}(\vec r)&=& i <\tilde{0}|[\hat{Q}_{sk},{\hat s}_s^{\rho}]|\tilde{0}> =
-4 \sum_{ij}^{K_i , K_j>0}
<ij|\hat{Q}_{sk}|\tilde 0>  \; \Im\{<ij|\hat{s}_s^{\rho}]|\tilde 0>\}
\nonumber\\
&=& 4 \sum_{ij}^{K_i , K_j >0}
\overline{<ij|\hat{Q}_{sk}|\tilde 0>} \;
s^s_{ij;\,\,\rho}(\rho,z)  \sin \mu\theta
= s_{Y, sk}^{\rho}(\rho,z) \sin \mu\theta ,
\label{eq:resp_s_rho_qp}\\
s_{Y, sk}^{z}(\vec r)&=& i <\tilde{0}|[\hat{Q}_{sk},{\hat s}_s^{z}]|\tilde{0}> =
-4 \sum_{ij}^{K_i , K_j>0}
<ij|\hat{Q}_{sk}|\tilde 0>  \; \Im\{<ij|\hat{s}_s^{z}]|\tilde 0>\}
\nonumber\\
&=& 4 \sum_{ij}^{K_i , K_j >0}
\overline{<ij|\hat{Q}_{sk}|\tilde 0>} \;
s^s_{ij;\,\,z}(\rho,z)  \sin \mu\theta
= s_{Y, sk}^{z}(\rho,z) \sin \mu\theta ,
\label{eq:resp_s_z_qp}\\
s_{Y, sk}^{\theta}(\vec r)&=& i <\tilde{0}|[\hat{Q}_{sk},{\hat s}_s^{\theta}]|\tilde{0}> =
2i \sum_{ij}^{K_i , K_j>0}
<ij|\hat{Q}_{sk}|\tilde 0>  2i \; \Im\{<ij|\hat{s}_s^{\theta}]|\tilde 0>\}
\nonumber\\
&=& -4 \sum_{ij}^{K_i , K_j >0}
\overline{<ij|\hat{Q}_{sk}|\tilde 0>} \;
s^s_{ij;\,\,\theta}(\rho,z) \cos \mu\theta
= s_{Y, sk}^{\theta}(\rho,z) \cos \mu\theta \; .
\label{eq:resp_s_theta_qp}
\end{eqnarray}
The time-even responses and their components are
\begin{eqnarray}
\rho_{X, sk}(\vec r)&=& i <\tilde{0}|[\hat{P}_{sk},{\hat \rho}_s]|\tilde{0}> =
-4i \sum_{ij}^{K_i ,K_j>0}
<ij|[\hat{P}_{sk}|\tilde 0>  \Re\{<ij|\hat{\rho}_s]|\tilde 0>\}
\nonumber\\
&=& 4 \sum_{ij}^{K_i , K_j >0}
  \overline{<ij|\hat{P}_{sk}|\tilde 0>} \;
 \rho^s_{ij}(\rho,z) \cos \mu\theta
= \rho_{X, sk}(\rho,z) \cos \mu\theta ,
\label{eq:resp_rho_qp}\\
\tau_{X, sk}(\vec r)&=& i <\tilde{0}|[\hat{P}_{sk},{\hat \tau}_s]|\tilde{0}> =
-4 \sum_{ij}^{K_i ,K_j>0}
<ij|[\hat{P}_{sk}|\tilde 0>  \Re\{<ij|\hat{\tau}_s]|\tilde 0>\}
\nonumber\\
&=& 4 \sum_{ij}^{K_i , K_j >0}
  \overline{<ij|\hat{P}_{sk}|\tilde 0>} \;
 \tau^s_{ij}(\rho,z) \cos \mu\theta
= \tau_{X, sk}(\rho,z) \cos \mu\theta ,
\label{eq:resp_tau_qp}\\
\Im_{Y, sk}^{\rho}(\vec r)&=& i <\tilde{0}|[\hat{P}_{sk},{\hat \Im}_s^{\rho}]|\tilde{0}> =
-4i \sum_{ij}^{K_i , K_j>0}
<ij|[\hat{P}_{sk}|\tilde 0> \Re\{<ij|\hat{\Im}_s^{\rho}]|\tilde 0>\}
\nonumber\\
&=& 4 \sum_{ij}^{K_i , K_j >0}
 \overline{<ij|\hat{P}_{sk}|\tilde 0>} \;
\Im^s_{ij;\,\,\rho}(\rho,z) \cos \mu\theta
= \Im_{X, sk}^{\rho}(\rho,z) \cos \mu\theta ,
\label{eq:resp_J_rho_qp}\\
\Im_{Y, sk}^{z}(\vec r)&=& i <\tilde{0}|[\hat{P}_{sk},{\hat \Im}_s^{z}]|\tilde{0}> =
-4i \sum_{ij}^{K_i , K_j>0}
<ij|[\hat{P}_{sk}|\tilde 0> \Re\{<ij|\hat{\Im}_s^{z}]|\tilde 0>\}
\nonumber\\
&=& 4 \sum_{ij}^{K_i , K_j >0}
\overline{<ij|\hat{P}_{sk}|\tilde 0>} \;
\Im^s_{ij;\,\, z}(\rho,z) \cos \mu\theta
= \Im_{X, sk}^{z}(\rho,z) \cos \mu\theta ,
\label{eq:resp_J_z_qp}\\
\Im_{Y, sk}^{\theta}(\vec r)&=& i <\tilde{0}|[\hat{P}_{sk},{\hat \Im}_s^{\theta}]|\tilde{0}> =
-4i \sum_{ij}^{K_i , K_j>0}
<ij|[\hat{P}_{sk}|\tilde 0> \Re\{<ij|\hat{\Im}_s^{\theta}]|\tilde 0>\}
\nonumber\\
&=& -4 \sum_{ij}^{K_i , K_j >0}
\overline{<ij|\hat{P}_{sk}|\tilde 0>} \;
\Im^s_{ij;\,\,\theta}(\rho,z) \; \sin \mu\theta
= \Im_{X, sk}^{\theta}(\rho,z) \sin \mu\theta .
\label{eq:resp_J_theta_qp}
\end{eqnarray}
The overline matrix elements are defined in (\ref{eq:over_P,X})-(\ref{eq:over_Q,Y}).
The pairing factors (\ref{pairing_factors}) are included into the matrix elements
and transition densities.
The values (\ref{vec_j_Y})-(\ref{eq:resp_J_theta_qp}) are
real. Time-even responses have diagonal contributions
$i=j$ while time-odd ones not.

For $\mu=0$, the responses fulfill (\ref{eq:res=0_mu=0}) and
\\
\\
\fbox{
\parbox{16cm}{
\begin{eqnarray} \label{eq:rot_resp}
 (\nabla \times \vec{j}_{Y, sk})_{\rho} &=& 0 ,
 \quad
  (\nabla \times \vec{j}_{Y, sk})_ {z} = 0 ,
\quad
 (\nabla \times \vec{j}_{Y, sk})_{\vartheta} =
  \partial_z j_{Y, sk}^{\rho} - \partial_{\rho} j_{Y, sk}^{z} ,
\nonumber\\
 (\nabla \times \vec{s}_{Y, sk})_{\rho} &=&
          -  \partial_z s_{Y, sk}^{\vartheta} ,
\quad
 (\nabla \times \vec{s}_{Y, sk})_{z} =
  \frac{1}{\rho} \partial_{\rho} (\rho s_{Y, sk}^{\vartheta}) ,
\quad
  (\nabla \times \vec{s}_{Y, sk})_{\vartheta} = 0 ,
\\
 (\nabla \times \vec{\Im}_{X, sk})_{\rho} &=& 0,
\quad
 (\nabla \times \vec{\Im}_{X, sk})_{z} = 0 ,
\quad
  (\nabla \times \vec{\Im}_{X, sk})_{\vartheta} =
   \partial_z \Im_{X, sk}^{\rho} - \partial_{\rho} \Im_{X, sk}^{z} .
\nonumber
\end{eqnarray}
}}
\\
\\
It is easy to see that the properties
(\ref{eq:res=0_mu=0})  and (\ref{eq:rot_resp}) of the responses
fully repeat the properties (\ref{eq:comb_Im_j=0})-(\ref{eq:comb_s=0}) and
(\ref{eq:rot_current}) of the
combinations of the transition densities, entering the responses.

\section{Coulomb contribution}
\label{sec:coul_inter}

The contribution of the Coulomb integral to the SRPA response is
\begin{equation}\label{Coul_resp}
U^{Coul}_k({\vec r})= e^2\int d\vec{r}\; '\frac{\rho_{X, pk}(\rho',z')
\cos\mu\theta'}{|\vec{r}-\vec{r}\; '|} .
\end{equation}
The typical trouble is  connected with the singularity at the point
$\vec{r}=\vec{r}'$. Because of the angular dependence $\cos\mu\theta'$ in the
nominator of (\ref{Coul_resp}), this trouble cannot be circumvented by common
methods.  For example, the  FTT solver method \cite{FFT} and procedures
\cite{Vauter_73,Stoitsov_Dob} do not assume
any angular dependence of this kind. So, we should develop our
own prescription.

First, we bypass the logarithmic
singularity by using the Vauterin identity \cite{Vauter_73,Stoitsov_Dob}
\begin{equation}\label{Vauterin relation}
 \Delta_{\vec{r}\; '}|\vec{r}-\vec{r}\; '|=\frac{2}{|\vec{r}-\vec{r}\; '|}.
\end{equation}
Then, after integration by parts, the integral (\ref{Coul_resp})
is reduced to
\begin{eqnarray}\label{Coul_2}
U^{Coul}_k (\vec{r}) &=& \frac{e^2}{2}\int d\vec{r}'|\vec{r}-\vec{r}\; '|
\Delta_{\vec{r}'}\rho_{X, pk}(\rho',z')\cos\mu\theta'
\nonumber \\
&=& \frac{e^2}{2} \int \rho' d\rho' dz'
[\partial^2_{\rho'} + \frac{1}{\rho'} \partial_{\rho'}
- \frac{\mu^2}{(\rho')^2} + \partial^2_{z'}]
\rho_{X, pk}(\rho',z')
\nonumber \\
&\cdot& \int_0^{2\pi} d\theta' |\vec{r}-\vec{r}\; '|\cos\mu\theta' .
\end{eqnarray}

For $\mu=0$, the angular integral in (\ref{Coul_2}) is recast to
\begin{equation}\label{ang_elliptic}
I (\vec{r};\rho',z') = \int_{0}^{2\pi}  d\theta'|\vec{r}-\vec{r}'|\cos\mu\theta' =
 4 \sqrt{d(\rho,z)} E(\frac{4\rho\rho'}{d(\rho,z)})
\end{equation}
where $d(\rho,z)=(\rho + \rho')^2 + (z-z')^2$ and
$E(\frac{4\rho\rho'}{d(\rho,z)})$ is the elliptic integral of the second order.
This integral can be approximated by a standard polynomial formula.
However, in the general case $\mu \ge 0$ we cannot get this result.
So, we should develop another procedure.

First of all, we should take into account that all the terms in (\ref{U_X_k})
have the common angular dependence
$\cos\mu\theta$. This should be the case for the Coulomb term as well.
To prove this, it is convenient to rewrite the angular integral from (\ref{Coul_2})
in the form
\begin{equation}\label{Coul_3}
I (\vec{r};\rho',z') = \int_{0}^{2\pi}  d\theta'|\vec{r}-\vec{r}\; '|\cos\mu\theta'
= I_1 \cos\mu\theta - I_2 \sin \mu\theta
\end{equation}
where
\begin{equation}
  I_1 (\rho,z;\rho',z') = \int_{0}^{2\pi} d\theta'|\vec{r}-\vec{r}\; '|\cos\mu(\theta'-\theta) ,
\quad
  I_2 (\rho,z;\rho',z') = \int_{0}^{2\pi} d\theta'|\vec{r}-\vec{r}'\; |\sin\mu(\theta'-\theta) .
\end{equation}
One may show that $I_2$=0 and $I_1$ does not depend on the angle $\theta$.
Hence (\ref{Coul_3}) is reduced to
\begin{equation}\label{Coul_5}
I = I_1 \cos\mu\theta
\end{equation}
thus delivering the desirable angular dependence in $(\ref{Coul_resp})$.

Altogether, the Coulomb contribution (\ref{Coul_resp}) to the response is recast
to the form
\begin{equation}\label{Coul_resp_recasted}
U^{Coul}_k({\vec r})= U^{Coul}_k (\rho,z) \cos\mu\theta
\end{equation}
with
\begin{eqnarray}\label{coul_rho_z}
U^{Coul}_k (\rho,z) &=&
e^2 \int \rho' d\rho' dz' \; I_1(\rho,z;\rho',z') \cdot
[\partial^2_{\rho'} + \frac{1}{\rho'} \partial_{\rho'}
- \frac{\mu^2}{(\rho')^2} + \partial^2_{z'}] \;
\rho_{X, pk}(\rho',z')
\nonumber \\
&=&
e^2 \int \rho' d\rho' dz'
\int_{0}^{\pi} d\theta_1
[(\rho + \rho')^2 + (z-z')^2-4\rho\rho'\cos^2 \theta_1]^{1/2}
\cos 2\mu\theta_1
\nonumber \\
&&\cdot
[\partial^2_{\rho'} + \frac{1}{\rho'} \partial_{\rho'}
- \frac{\mu^2}{(\rho')^2} + \partial^2_{z'}] \;
\rho_{X, pk}(\rho',z') \; .
\end{eqnarray}
The matrix element (\ref{eq:<iX|j>_coul}) reads as
\begin{equation}\label{eq:X_ij_coul}
 <i|\hat{X}_{pk}^{p}|j>_{Coul} =
 e^2 \int \rho d\rho dz \; U^{Coul}_k(\rho,z)
\rho_{ij}^{s}(\rho,z) \cdot
\int_0^{2\pi} d\theta  \cos \mu\theta e^{-i\mu\theta} \; .
\end{equation}

\section{Matrix elements}
\label{sec:matr_el}

This Appendix represents the matrix elements of the operators $\hat P_{sk}$,
$\hat X^{s'}_{sk}$ and $\hat Y^{s'}_{sk}$. The pairing factors are supposed to
be included into the matrix elements.

\subsection{Matrix elements of operator $\hat P_{sk}$}
\label{sec:matr_el_P}

The matrix element of operator $\hat P_{sk}$ reads
\begin{eqnarray} \label{me_P_ij}
<ij|P_{sk}|\tilde 0> &=&
 2i \epsilon_{ij}
<ij|Q_{sk}|\tilde 0> - <ij|Y_{sk}^s|\tilde 0>
\\
&=& i
\{
(2 \epsilon_{ij} \overline{<ij|Q_{sk}|\tilde 0>} -
\overline{<ij|Y^{s(1)}_{sk}|\tilde 0>}) \;
 (\delta_{\mu, K_i - K_j} + \delta_{-\mu, K_i - K_j})
\nonumber\\
 && \qquad\qquad \qquad +
 \overline{<ij|Y^{s(2)}_{sk}|\tilde 0>} \;
(\delta_{\mu, K_i - K_j}-\delta_{\mu, K_i - K_j})
\}
\nonumber
\end{eqnarray}
for $\mu=|K_i - K_j|$ and
\begin{eqnarray}\label{me_P_ibarj}
<i\bar j|P_{sk}|\tilde 0> &=&
 2i \epsilon_{ij} <i\bar j|Q_{sk}|\tilde 0>  - <i\bar j|Y_{sk}^s|\tilde 0>
\\
&=& i
\{
2 \epsilon_{ij}  \overline{<i\bar j|Q_{sk}|\tilde 0>}
- \overline{<i\bar j|Y^{s(1)}_{sk}|\tilde 0>}
+ \overline{<i\bar j|Y^{s(2)}_{sk}|\tilde 0>} \}
\delta_{\mu, K_i + K_j}
\nonumber
\end{eqnarray}
for $\mu=K_i + K_j$. Here, $K_i$ and $K_j$ are projections of the complete single-particle
moment onto symmetry z-axis of the axial nucleus.
Expressions for $<ij|Y^{s(1)}_{sk}|\tilde 0>$
and $<ij|Y^{s(2)}_{sk}|\tilde 0>$ are given in the section \ref{sec:matr_el_Y}.
The combinations of Kronecker symbols follow from the angular integrals.
Because of the specific combinations in (\ref{me_P_ij}),
all the terms of this matrix element
have the same permutation properties for $i \leftrightarrow j$.

\subsection{Matrix elements of operator $\hat X^{s'}_{sk}$}
\label{sec:matr_el_X}

The matrix element of operator $\hat X^{s'}_{sk}$ reads
\begin{eqnarray} \label{X_matr_el}
<ij|\hat{X}_{sk}^{s'}|\tilde 0>
&=& \overline{<ij|\hat{X}_{sk}^{s'(1)}|\tilde 0>}(\delta_{\mu,K_i - K_j}+\delta_{\mu,K_j - K_i})
\\
&+& \overline{<ij|\hat{X}_{sk}^{s'(2)}|\tilde 0>}(\delta_{\mu,K_i - K_j}-\delta_{\mu,K_j - K_i}) ,
\nonumber
\end{eqnarray}
for $\mu=|K_i-K_j|$ and
\begin{equation} \label{X_matr_el_ibarj}
<i\bar j|\hat{X}_{sk}^{s'}|\tilde 0>
=
\{
\overline{<i\bar j|\hat{X}_{sk}^{s'(1)}|\tilde 0>}
+ \overline{<i\bar j|\hat{X}_{sk}^{s'(2)}|\tilde 0>}
\}\delta_{\mu,K_i + K_j}
\;
\end{equation}
for $\mu=K_i+K_j$.

Here
\begin{eqnarray}
\overline{<ij|\hat{X}_{sk}^{s'(1)}|\tilde 0>} &=& \pi
\int \rho d\rho dz \;
[ {\bar U}_{sk}^{s'}(\rho,z) \rho^{s'}_{ij}(\rho,z)
+ {\bar B}_{sk}^{s'}(\rho,z)\tau^{s'}_{ij}(\rho,z)
\\
\label{X_1_matr_el}
&+& {\bar W}_{sk; \rho}^{s'}(\rho,z)\Im^{s'}_{ij;\rho}(\rho,z)
+ {\bar W}_{sk; z}^{s'}(\rho,z)\Im^{s'}_{ij;z}(\rho,z) ]
\nonumber\\
&+& \delta_{s,s'}\overline{<ij|\hat{X}_{sk}^{s'(pair)}|\tilde 0>}
\nonumber\\
\label{X_2_matr_el}
\overline{<ij|\hat{X}_{sk}^{s'(2)}|\tilde 0>} &=& \pi
\int \rho d\rho dz \; {\bar W}_{sk;\theta}^{s'}(\rho,z)\Im^{s'}_{ij;\theta}(\rho,z) \; ,
\\
\overline{<ij|\hat{X}_{sk}^{s(pair)}|\tilde 0>}
&=& - \delta_{\mu,0}  \delta_{i,j} G_s \chi_{X,\; ks} (u^2_i -v^2_i) \; .
\label{eq:pair_X_ij_me}
\end{eqnarray}
The values entering (\ref{eq:pair_X_ij_me}) are defined in Sec. \ref{sec:pairing}.

Further
\begin{eqnarray}\label{bar_U_k}
{\bar U}_{sk}^{s'}(\rho,z)
&=&
\{
 b_0-b_0'\delta_{s,s'}
\\
 &+& [-b_2+b_2'\delta_{s,s'}]
[\partial^2_{\rho}+\frac{1}{\rho}\partial_{\rho}-\frac{\mu^2}{\rho^2}+\partial^2_z]
+b_3\frac{(\alpha +1)(\alpha +2)}{3}\rho^{\alpha}(\rho,z)
\nonumber\\
&-&
b_3'[\frac{\alpha(\alpha -1)}{3}\rho^{\alpha -2}(\rho,z)
\sum_{s"}\rho^2_{s"}(\rho,z)
\nonumber\\
&+&
\frac{2}{3}\alpha \rho^{\alpha -1}(\rho,z)
[\rho_{s}(\rho,z)+\rho_{s'}(\rho,z)]+\frac{2}{3}\rho^{\alpha}(\rho,z)\delta_{s,s'}]
\nonumber\\
&-&\delta_{s',p}\delta_{s,p}\frac{e^2}{3}(\frac{3}{\pi})^{1/3}\rho_p^{-2/3}(\rho,z)
\}
\; \rho_{X, sk}(\rho,z)
+ \delta_{s',p}\delta_{s,p} U_{k}^{(\rm Coul)}(\rho,z)
\nonumber\\
&+&[b_1-b_1'\delta_{s,s'}] \; \tau_{X, sk}(\rho,z)
\nonumber\\
&-&[b_4+b_4'\delta_{s,s'}]
[\frac{1}{\rho} \partial_{\rho} (\rho \Im^{\rho}_{X, sk}(\rho,z))
+\partial_z \Im^{z}_{X, sk}(\rho,z))
+\frac{\mu}{\rho} \Im^{\theta}_{X, sk}(\rho,z)]
\;,
\nonumber
\end{eqnarray}
\begin{eqnarray}
\label{bar_B_k}
{\bar B}_{sk}^{s'}(\rho,z) &=&
[b_1-b_1'\delta_{s, s'}]\rho_{X, sk}(\rho,z)\; ,
\\
\label{bar_W_k_rho}
{\bar W}_{sk;\;\rho}^{s'}(\rho,z)&=&
 [b_4+b_4'\delta_{s, s'}]\partial_{\rho}\rho_{X, sk}(\rho,z)\: ,
\\
\label{bar_W_k_z}
{\bar W}_{sk;\;z}^{s'}(\rho,z)&=&
[b_4+b_4'\delta_{s,s'}]\partial_{z}\rho_{X, sk}(\rho,z)\: ,
\\
\label{bar_W_k_theta}
{\bar W}_{sk;\;\theta}^{s'}(\rho,z)&=&
- [b_4+b_4'\delta_{s,s'}]\frac{\mu}{\rho} \rho_{X, sk}(\rho,z)\: .
\end{eqnarray}
The overline matrix elements are real.
The combinations of Kronecker symbols follow from the angular integrals.
Expressions for $U_{k}^{(\rm Coul)}(\rho,z)$ in
(\ref{bar_U_k}) is done in (\ref{coul_rho_z}).
The matrix elements (\ref{X_matr_el_ibarj}) and
(\ref{X_2_matr_el}) vanish at $\mu=0$. Instead, the pairing matrix element
(\ref{eq:pair_X_ij_me}) exists only  at $\mu=0$.

\subsection{Matrix elements of operator $\hat Y^{s'}_{sk}$}
\label{sec:matr_el_Y}.

The matrix element of operator $\hat Y^{s'}_{sk}$ reads
\begin{eqnarray}\label{Y_matr_el_gen}
 <ij|\hat{Y}_{s k}^{s'}|\tilde 0> &=&
i\{
\overline{<ij|\hat{Y}_{s k}^{s'(1)}|\tilde 0>}(\delta_{\mu,K_i - K_j}+\delta_{\mu,K_j - K_i})
\nonumber \\
&-& \overline{<ij|\hat{Y}_{s k}^{s'(2)}|\tilde 0>}(\delta_{\mu,K_i - K_j}-\delta_{\mu,K_j - K_i})
\} ,
\label{eq:Y_over_me}
\end{eqnarray}
for $\mu=|K_i-K_j|$ and
\begin{equation} \label{Y_matr_el_ibarj}
<i\bar j|\hat{Y}_{sk}^{s'}|\tilde 0> = i
\{
\overline{<i\bar j|\hat{Y}_{sk}^{s'(1)}|\tilde 0>}
- \overline{<i|\hat{Y}_{sk}^{s'(2)}|j>}
\}\delta_{\mu,K_i + K_j}
\end{equation}
for $\mu=K_i+K_j$. The overline matrix elements are real.

Further
\begin{eqnarray}\label{Y_1_matr_el}
\overline{<ij|\hat{Y}_{sk}^{s'(1)}|\tilde 0>} &=& \pi \int \rho d\rho dz
\\
&\cdot&[{\bar A}_{sk; \rho}^{s'}(\rho,z) j^{s'}_{ij; \rho}(\rho,z)
+{\bar A}_{sk; z}^{s'}(\rho,z) j^{s'}_{ij; z}(\rho,z)
+{\bar S}_{sk; \theta}^{s'}(\rho,z) s^{s'}_{ij; \theta}(\rho,z)] ,
\nonumber\\
\label{Y_2_matr_el}
\overline{<ij|\hat{Y}_{sk}^{s'(2)}|\tilde 0>} &=& \pi \int \rho d\rho dz
\\
&\cdot&
[{\bar A}_{sk; \theta}^{s'}(\rho,z) j^{s'}_{ij; \theta}(\rho,z)
+{\bar S}_{sk; \rho}^{s'}(\rho,z) s^{s'}_{ij; \rho}(\rho,z)
+{\bar S}_{sk; z}^{s'}(\rho,z) s^{s'}_{ij; z}(\rho,z)] .
\nonumber
 \end{eqnarray}
and
\begin{eqnarray}
{\bar A}_{sk; \rho}^{s'}(\rho,z)&=&
 2 [- b_1 + b_1' \delta_{s,s'}] j_{Y, sk; \rho}(\rho,z)
\\ \nonumber
&-& [b_4 + b_4' \delta_{s,s'}]
[\frac{\mu}{\rho}s_{Y, sk; z}(\rho,z)-\partial_z s_{Y, sk; \theta}(\rho,z)] \:,
\\
{\bar A}_{sk; z}^{s'}(\rho,z)&=&
2 [- b_1 + b_1' \delta_{s,s'}] j_{X, sk; z}
\\ \nonumber
&-& [b_4 + b_4' \delta_{s,s'}]
[\frac{1}{\rho}\partial_{\rho}(\rho s_{Y, sk; \theta})-\frac{\mu}{\rho}s_{Y, sk; \rho}]
\\
{\bar A}_{sk; \theta}^{s'}(\rho,z)&=&
2 [- b_1 + b_1' \delta_{s,s'}] j_{Y, sk; \theta}
\\ \nonumber
&-& [b_4 + b_4' \delta_{s,s'}]
[\partial_z s_{Y, sk; \rho}-\partial_{\rho} s_{Y, sk; z}] \:,
\\
{\bar S}_{sk; \rho}^{s'}(\rho,z)&=&
 [b_4 + b_4' \delta_{s,s'}]
[\frac{\mu}{\rho}j_{Y, sk; z}(\rho,z)+\partial_z j_{Y, sk; \theta}(\rho,z)] \:,
\\
{\bar S}_{sk; z}^{s'}(\rho,z)&=&
 -[b_4 + b_4' \delta_{s,s'}]
[\frac{1}{\rho}\partial_{\rho}(\rho j_{Y, sk; \theta})+\frac{\mu}{\rho}j_{Y, sk; \rho}] \:,
\\
{\bar S}_{sk; \theta}^{s'}(\rho,z)&=&
  -[b_4 + b_4' \delta_{s,s'}]
[\partial_z j_{Y, sk; \rho}-\partial_{\rho} j_{Y, sk; z}] \;.
\end{eqnarray}
For $\mu =0$, one puts
\begin{equation}
{\bar A}_{sk; \theta}^{s'}(\rho,z)={\bar S}_{sk; \rho}^{s'}(\rho,z)
={\bar S}_{sk; z}^{s'}(\rho,z) =0
\end{equation}
since these terms  adjoin $\sin \mu\theta$.
The remaining terms are reduced to
\begin{eqnarray}
{\bar A}_{sk; \rho}^{s'}(\rho,z)&=&
 2 [- b_1 + b_1' \delta_{s,s'}] j_{Y, sk; \rho}(\rho,z)
- [b_4 + b_4' \delta_{s,s'}] \partial_z s_{Y, sk; \theta}(\rho,z) \:,
\\
{\bar A}_{sk; z}^{s'}(\rho,z)&=&
2 [- b_1 + b_1' \delta_{s,s'}] j_{Y, sk; z}
- [b_4 + b_4' \delta_{s,s'}]
\frac{1}{\rho}\partial_{\rho}(\rho s_{Y, sk; \theta}) \: ,
\\
{\bar S}_{sk; \theta}^{s'}(\rho,z)&=&
  -[b_4 + b_4' \delta_{s,s'}]
[\partial_z j_{Y, sk; \rho}-\partial_{\rho} j_{Y, sk; z}] \;.
\end{eqnarray}

\end{document}